\documentclass[12pt]{amsart}
\usepackage{amssymb}
\usepackage{hyperref}

\newtheorem{thm}{Theorem}[section]

\newtheorem{cor}[thm]{Corollary}
\newtheorem{lem}[thm]{Lemma}
\newtheorem{prop}[thm]{Proposition}

\newtheorem*{thm*}{Theorem}
\newtheorem*{claim1}{Claim I}
\newtheorem*{claim2}{Claim II}
\newtheorem*{claim3}{Claim III}
\newtheorem*{claim4}{Claim IV}

\theoremstyle{definition}
\newtheorem{defn}[thm]{Definition}
\newtheorem*{defn*}{Definition}
\newtheorem{rem}[thm]{Remark}
\numberwithin{equation}{section}

\newcommand{\al}{\alpha}
\newcommand{\be}{\beta}
\newcommand{\ga}{\gamma}
\newcommand{\Ga}{\Gamma}
\newcommand{\de}{\delta}
\newcommand{\la}{\lambda}
\newcommand{\La}{\Lambda}
\newcommand{\om}{\omega}
\newcommand{\Om}{\Omega}
\newcommand{\si}{\sigma}
\newcommand{\ka}{\kappa}
\newcommand{\eps}{\varepsilon}

\newcommand{\C}{\mathbb C}
\newcommand{\R}{\mathbb R}
\newcommand{\Y}{\mathbb Y}
\newcommand{\Z}{\mathbb Z}

\newcommand{\wt}{\widetilde}

\newcommand{\A}{\mathcal{A}}
\newcommand{\X}{\mathfrak{X}}

\DeclareMathOperator{\CONF}{Conf}
\DeclareMathOperator{\RES}{Res}
\DeclareMathOperator{\JUMP}{Jump}
\DeclareMathOperator{\const}{const}
\DeclareMathOperator{\gam}{GAMMA}
\DeclareMathOperator{\sgn}{sgn}

\newcommand{\re}{\mathop{\mathrm{Re}}}
\newcommand{\im}{\mathop{\mathrm{Im}}}
\newcommand{\Tr}{\mathop{\mathrm{Tr}}}

\newcommand{\bp}{\mathbf p}

\begin{document}

\title
 {Giambelli compatible point processes}

\author{Alexei Borodin}
\address{Department of Mathematics,  253-37,  Caltech, Pasadena, CA 91125}
\email{borodin@caltech.edu}

\author{Grigori Olshanski}
\address{Institute for Information Transmission Problems,
Bolshoy Karetny 19, Moscow 127994, GSP-4, Russia}
\email{olsh@online.ru}

\author{Eugene Strahov}
\address{Department of Mathematics,  253-37,  Caltech, Pasadena, CA 91125}
\email{strahov@caltech.edu}

\thanks{The present research was partially conducted during the period  the
first named author (A.~B.) served as a Clay Mathematics Institute Research
Fellow; he was also partially supported by the NSF grant DMS-0402047 and the
CRDF grant RIM1-2622-ST-04. The second named author (G.~O.) was supported by
the CRDF grant RM1-2543-MO-03.}

\dedicatory{Dedicated to Amitai Regev on the occasion of his
65th birthday}
\begin{abstract}
We distinguish a class of random point processes which we call
Giambelli compatible point processes. Our definition was partly
inspired by determinantal identities for averages of products and
ratios of characteristic polynomials for random matrices found
earlier by Fyodorov and Strahov. It is closely related to the
classical Giambelli formula for Schur symmetric functions.

We show that orthogonal polynomial ensembles, z-measures on
partitions, and spectral measures of characters of generalized
regular representations of the infinite symmetric group generate
Giambelli compatible point processes. In particular, we prove
determinantal identities for averages of analogs of characteristic
polynomials for partitions.

Our approach provides a direct derivation of determinantal
formulas for correlation functions.

\end{abstract}
\maketitle

\section*{Introduction}

This paper appeared as a result of our attempt to find a
connection between the work of Fyodorov and Strahov on
evaluating the averages of products and ratios of characteristic
polynomials of random matrices, and measures on partitions which
exhibit random matrix type behavior.

Among many other things, Fyodorov and Strahov \cite{FS},
\cite{SF} proved the following formula: Let $H$ be a random
Hermitian $N\times N$ matrix distributed according to the
Gaussian measure
$$
P(dH)=\const\cdot \exp(-\Tr(H^2))dH
$$
and $D(z)=\det(z-H)$ be
its characteristic polynomial. Then for any $d=1,2,\dots$ and
$u_1,\dots,u_d\in \C\setminus \R$, $v_1,\dots,v_d\in\C$,
\begin{equation}\label{A}
\left\langle\frac{D(v_1)\cdots D(v_d)}{D(u_1)\cdots
D(u_d)}\right\rangle=\det\left(\frac 1{u_i-v_j}\right)^{-1}\cdot
\det\left\langle\frac1{u_i-v_j}\,\frac{D(v_j)}{D(u_i)}\right\rangle
\end{equation}
(both determinants have size $d\times d$). By now this result
has a number of different proofs and extensions, see \cite{FS},
\cite{SF}, \cite{BH1}, \cite{BH2}, \cite{BH3}, \cite{BH4},
\cite{BDS}, \cite{BS}, \cite{AP}, \cite{BG}, \cite{CFS},
\cite{GGK}. Formulas of this type are of interest in quantum
physics and classical number theory, see \cite{AS}, \cite{Fyo},
\cite{CFKRS}, \cite{HKO}, \cite{KS1}, \cite{KS2}, \cite{KS3}.
Apart from that, (\ref{A}) provides a convenient way to show
that the correlation functions of the eigenvalues of $H$ can be
written as determinants of a certain kernel. \footnote{This
basic fact lies at the foundation of Random Matrix Theory, see
e.g. \cite{Me}, \S5.2.}

On the other hand, in recent years there has been a considerable
interest in measures on partitions which are in many ways
similar to the eigenvalue distributions in Random Matrix Theory.
The sources of such measures are quite diverse; they include
combinatorics, representation theory, random growth models,
random tilings, etc. In this paper we concentrate on the
so-called $z$-measures which arise naturally in representation
theory of the infinite symmetric group. This 3-parameter family
of measures contains a number of other interesting measures on
partitions (including the Plancherel measures and measures
arising in last passage percolation models) as degenerations,
see \cite{BO-RSK}.

One natural question is: What is the analog of formula (\ref{A})
for random partitions? Note that the very existence of such an
analog is rather nontrivial: it is not {\it a priori\/} clear
what a ``characteristic polynomial of a partition'' is, and the
finite-dimensional averaging in (\ref{A}) should be replaced by
essentially an infinite-dimensional one over the space of all
partitions.

The main goal of this paper is to provide an analog of (\ref{A})
for the $z$-measures on partitions and their representation
theoretic scaling limits, explain a general mechanism of where
the identities of type (\ref{A}) come from, and show how these
identities imply the determinantal structure of the correlation
functions of the underlying point processes. Remarkably, this
approach provides the most straightforward derivation of the
associated correlation kernels among those known so far.

Let us proceed to a more detailed description of the content of
the paper.

\subsection*{ a) Giambelli compatible processes}

Let us first introduce some notation. Let $\Lambda$ be the
algebra of symmetric functions and $\{s_\lambda\}$ be its basis
consisting of the Schur functions. The Schur functions are
parameterized by partitions $\lambda=(\lambda_1\ge \lambda_2\ge
\dots\ge 0)$ which can also be written in the Frobenius
notation:
$$
\lambda=(p_1,\dots,p_d\mid q_1,\dots,q_d)
$$
(see \S1 or \cite{Ma} for definitions). The Schur functions
satisfy the following basic identity called the {\em Giambelli
formula\/}:
$$
s_{(p_1,\dots,p_d\mid q_1,\dots,q_d)}=\det \bigl[s_{(p_i\mid
q_j)}\bigr]_{i,j=1}^d\,.
$$

It turns out that the following remarkable fact holds true:
Denote by $\langle s_\lambda \rangle$ the average of the Schur
function $s_\lambda$ evaluated at the eigenvalues of matrix $H$
with respect to the Gaussian measure on $H$ introduced above.
Then for any partition $\lambda=(p_1,\dots,p_d\mid
q_1,\dots,q_d)$
\begin{equation}\label{B} \langle
s_{(p_1,\dots,p_d\mid q_1,\dots,q_d)}\rangle =\det \bigl[\langle
s_{(p_i\mid q_j)}\rangle \bigr]_{i,j=1}^d
\end{equation}
or, in other words, the Giambelli formula remains invariant
under the averaging.

This fact is closely related to the identity (\ref{A}). More
exactly, our first result is the following

\begin{claim1} Let $\langle\,\cdot\,\rangle$ be an arbitrary
linear map from $\Lambda$ to $\C$. Then the following two
conditions are equivalent:

{\rm (i)} For any $d=1,2,\dots$ and any integers
$p_1>\dots>p_d\ge 0$, $q_1>\dots>q_d\ge 0$, the averaged
Giambelli formula (\ref{B}) holds.

{\rm (ii)} For any $d=1,2,\dots$ the following formal power
series identity holds:
\begin{equation}\label{C}
\left\langle H(u_1)\cdots H(u_d)E(v_1)\cdots
E(v_d)\right\rangle=\det\left(\frac 1{u_i+v_j}\right)^{-1}\,
\det\left\langle\frac{H(u_i)E(v_i)}{u_i+v_j}\right\rangle
\end{equation}
where both determinants are of size $d\times d$, and $H(u)$ and
$E(v)$ are the generating functions of the one-row and
one-column Schur functions:
$$
H(u)=1+\sum\limits_{k=1}^{\infty}\frac{s_{(k)}}{u^k}\,,\;\;\;
E(v)=1+\sum\limits_{k=1}^{\infty} \frac{s_{(1^k)}}{v^k}\,.
$$
\end{claim1}

If we now evaluate the symmetric functions at $N$ eigenvalues
$x_1,\dots,x_N$ of $H$, then
$$
H(u)=\prod_{i=1}^N\frac 1{1-x_i u^{-1}},\qquad
E(v)=\prod_{i=1}^N\left(1+x_i v^{-1}\right),
$$
and averaging over $H$ turns (\ref{C}) into
(\ref{A}).\footnote{A careful reader might object that (\ref{C})
is a formal power series identity while (\ref{A}) is an identity
of actual functions in $u_i$'s and $v_j$'s. It does require some
efforts to pass from one to the other and this issue will be
addressed in the body of the paper.}

We also show that in condition (i) above the Schur functions may
be replaced by the multiparameter Schur functions (see \S3 in
\cite{ORV} or \S1.2 below) or by their special case --- the
Frobenius-Schur functions (see \S2 in \cite{ORV} and \S1.3).

The next definition is inspired by Claim I.

\begin{defn*} A random point process (= a probability
measure on  point configurations) is called {\it Giambelli
compatible\/} if there exists a homomorphism of the algebra of
symmetric functions $\Lambda$ to a suitable algebra of functions
on point configurations such that the linear functional on
$\Lambda$ obtained by averaging the images of symmetric
functions satisfies the conditions of Claim I.
\end{defn*}

In this terminology the point process of eigenvalues of random
Hermitian matrices with the Gaussian measure is Giambelli
compatible.

In this paper we discuss three examples of Giambelli compatible
random point processes. Let us describe them one by one.

\subsection*{ b) Orthogonal polynomial ensembles}
Let $\mu$ be an arbitrary measure on $\R$ with finite moments.
The $N$-point orthogonal polynomial ensemble on $\R$ associated
with $\mu$ is a probability measure on $\R^N$ of the form
$$
P_N(dx_1,\dots,dx_N)=\const\cdot \prod_{1\le i<j\le N}
(x_i-x_j)^2 \prod_{i=1}^N \mu(dx_i).
$$
Orthogonal polynomial ensembles are very common in Random Matrix
Theory; they are also often called ``$\beta=2$
ensembles''.\footnote{The value of $\beta$ refers to the power
of the Vandermonde determinant.} In particular, for any even
degree polynomial $V(x)$ with a nonnegative highest coefficient,
the radial part (= projection to eigenvalues) of the unitarily
invariant probability measure $$\const\cdot \exp(-\Tr(V(H)))
dH$$ on the Hermitian $N\times N$ matrices is an orthogonal
polynomial ensemble with $\mu(dx)=\exp(-V(x))dx$, see e.g.
\cite{De}. Orthogonal polynomial ensembles with discretely
supported measures $\mu$ are also quite popular, see e.g.
\cite{Jo-shape}, \cite{Jo-planch}, \cite{Jo-nonint},
\cite{BO-RSK}, \cite{BO-Uinfty}, \cite{BO-Gamma}, \cite{KOR},
\cite{O'C}.

\begin{claim2} Any orthogonal polynomial ensemble defines a
Giambelli compatible process with respect to the standard
realization of the symmetric functions as functions on $\R^N$.
\end{claim2}

This fact (more exactly, formula \ref{A}) allows one to derive
the determinantal formula for the correlation functions of the
orthogonal polynomial ensembles, and to express the correlation
kernel in terms of the 2-point average $\langle
H(u)E(v)\rangle$. This average is in its turn expressible
through the orthogonal polynomials associated with $\mu$. See
\S3 for details.

\subsection*{c) $z$-measures on partitions}
These probability measures depend on three (generally speaking,
complex) parameters $z,z',\xi$ and assign to a partition
$\lambda$ with Frobenius coordinates $(p_1,\dots,p_d\mid
q_1,\dots,q_d)$ the weight
\begin{multline}\label{D}
M_{z,z',\xi}(\lambda)=(1-\xi)^{zz'}
\xi^{|\lambda|}\\
\times (zz')^d\prod_{i=1}^d \frac{
(z+1)_{p_i}(z'+1)_{p_i}(-z+1)_{q_i}(-z'+1)_{q_i}}{(p_i!)^2(q_i!)^2}
\,{\det}^2\left(\frac{1}{p_i+q_j+1}\right).
\end{multline}
Here $(a)_k=a(a+1)\cdots (a+k-1)$ is the Pochhammer symbol.

There are various sets of conditions on $(z,z',\xi)$ that
guarantee that these weights are nonnegative and their sum over
all partitions is equal to 1; for instance, one can take
$z'=\bar{z}\in\C$ and $\xi\in (0,1)$.

The $z$-measures describe the generalized regular
representations of the infinite symmetric group. Briefly,
$M_{z,z',\xi}(\lambda)$ are essentially the Fourier coefficients
of characters of such representations. We refer to \cite{KOV},
\cite{Ol-surv} for details.

Apart from that, the $z$-measures degenerate to a variety of
measures of rather different origins. When both parameters $z$
and $z'$ are positive integers, the $z$-measures arise in a last
passage percolation model, see \cite{Jo-shape}, while when $z$
and $z'$ are integers of different signs, the corresponding
measures are directly related to the ``digital boiling'' growth
model, see \cite{GTW}. In the limit $z'\to \infty$, $\xi\to 0$,
and with integral $z\in\Z_+$ the $z$-measures are obtained from
pushforwards of the uniform measures on random words built out
of an alphabet with $z$ letters under the Robinson-Schensted
correspondence, see e.g. \cite{BO-RSK}. Finally, in the limit
when both $z$ and $z'$ tend to infinity and $\xi\to 0$, the
$z$-measure becomes the celebrated poissonized Plancherel
measure, see e.g. \cite{BOO}.

It is convenient to identify partitions
$\lambda=(p_1,\dots,p_d\mid q_1,\dots, q_d)$ with finite point
configurations on $\Z'=\Z+\frac 12$ as follows
\begin{equation}\label{E}
\lambda \longleftrightarrow \bigl\{-q_1-\tfrac 12,-q_2-\tfrac
12,\dots,-q_d-\tfrac 12,p_d+\tfrac 12,\dots, p_2+\tfrac 12,
p_1+\tfrac 12\bigr\}\subset \Z'.
\end{equation}
Then any measure on partitions, in particular, the $z$-measure,
defines a random point process on $\Z'$.

In order to move on, we need to realize the symmetric functions
as functions on partitions. A suitable for us way of doing that
was suggested in \cite{KO}. Namely, the Newton power sums
$\bp_k\in\La$ (do not confuse with Frobenius coordinates $p_i$)
are specialized as follows: For $\lambda=(p_1,\dots,p_d\mid
q_1,\dots, q_d)$
$$
\bp_k(\lambda)=\sum_{i=1}^d (p_i+\tfrac12)^k
+(-1)^{k-1}(q_i+\tfrac12)^k.
$$
Then the algebra $\Lambda$ is being mapped to the algebra of
{\it polynomial functions\/} on partitions. The images of the
generating series $H(u)$ and $E(v)$ under this map have the form
$$
\gathered H(u)(\lambda)=\prod\limits_{i\ge 1}\frac{u+i-\tfrac
12}{u-\lambda_i+i-\frac 12}=\prod\limits_{i\ge
1}\frac{u+\lambda_i'-i+\tfrac 12}{u-i+\frac
12}=\prod\limits_{i=1}^d\frac{u+q_i+\tfrac
12}{u-p_i-\frac 12}\,, \\
E(v)(\lambda)=\prod\limits_{i\ge1}\frac{v+\lambda_i-i+\frac 12}{v-i+\frac
12}=\prod\limits_{i\ge 1}\frac{v+i-\tfrac 12}{v-\lambda_i'+i-\frac
12}=\prod\limits_{i=1}^d\frac{v+p_i+\frac 12}{v-q_i-\frac 12}\,.
\endgathered
$$
These are the analogs of the characteristic polynomial and its
inverse for partitions (here $\lambda'$ denotes the transposed
partition).

\begin{claim3} The random point process on $\Z'$
corresponding to any $z$-measure or any of its degenerations is
Giambelli compatible with respect to the realization of the
algebra of symmetric functions on partitions described above.
\end{claim3}

While for the orthogonal polynomial ensembles the derivation of
the determinantal formula for the correlation functions from
Giambelli compatibility is of rather limited interest, for the
$z$-measures such a derivation provides the simplest known proof
of this important fact.

For any finite subset $X$ of $\Z'$ let us denote by $\rho(X)$
the $z$-measure probability that the random point configuration
(\ref{E}) contains $X$. Claim III leads to the following result.

\begin{thm*} For any finite set $X=\{x_1,\dots,x_m\}\subset
\Z'$ we have
$$
\rho(X)=\det[K(x_i,x_j)]_{i,j=1}^m,
$$
where
\begin{equation}\label{H}
K(x,y)=\begin{cases}
   \underset{u=y}{\RES} \;\dfrac{\langle H(u)E(-x)\rangle}{x-y},
   & x>0, y>0,\; x\neq y, \\
    -\underset{v=x}{\RES}\;\underset{u=y}{\RES}\;
    \dfrac{\langle H(u)E(-v)\rangle}{x-y}, &  x<0, y>0,\\
    \dfrac{\langle H(y)E(-x)\rangle}{x-y}, & x>0, y<0, \\
   -\underset{v=x}{\RES}\;\dfrac{\langle H(y)E(-v)\rangle}{x-y},
    & x<0, y<0,\; x\neq y.  \\
\end{cases}
\end{equation}
Here $\langle\,\cdot\,\rangle$ means averaging over the
$z$-measure $M_{z,z',\xi}$, and the indeterminacy arising for
$x=y$ is resolved via the L'Hospital rule.
\end{thm*}

It is now immediate to explicitly evaluate (using formula
\ref{D}) the 2-point average $\langle H(u)E(v)\rangle$ and the
whole correlation kernel $K(x,y)$ in terms of the Gauss
hypergeometric function. We do this simple computation in the
body of the paper and thus rederive the hypergeometric kernel of
\cite{BO-hyper}.\footnote{It should be noted that the proof in
\cite{BO-hyper} was not a derivation but a verification. Known
derivations of the hypergeometric kernel are somewhat indirect:
they use an $sl(2)$-action on the infinite wedge space
\cite{Ok-SL2}, more general {\it Schur} measures on partitions
\cite{Ok-infw}, \cite{BOk}, \cite{Jo-ECM}, \cite{Ra}, or
nontrivial analytic continuation arguments \cite{BO-dynam}.}
Details on $z$-measures on partitions are presented in \S4.

\subsection*{ d) Spectral $z$-measures}
These measures describe the (spectral) decomposition of the
generalized regular representations of the infinite symmetric
group on irreducibles. The spectral $z$-measures have continual
infinite-dimensional support and they are not easy to describe
in simple terms.

One way to obtain the spectral $z$-measures is to take a certain
scaling limit of the $z$-measures on partitions described above
as $\xi\to 1$. Another, more direct approach is to represent
them as a unique solution of an infinite-dimensional moment
problem.

More exactly, the probability measures that we are interested in
live on the space of pairs of nonincreasing sequences
$(\alpha,\beta)$ of nonnegative real numbers whose total sum is
finite:
$$
\alpha_1\ge \alpha_2\ge \dots\ge 0,\quad \beta_1\ge \beta_2\ge
\dots\ge 0,\qquad \sum_{i}\alpha_i+\sum_j \beta_j<\infty.
$$
There is a standard way of realizing the algebra of symmetric
functions $\Lambda$ by functions on such pairs of sequences.
Namely, the images of the Newton power sums take the form (cf.
(\ref{E}))
\begin{equation}\label{F}
\bp_k(\alpha,\beta)=\sum_{i}\alpha_i^k+(-1)^{k-1}\sum_j
\beta_j^k,
\end{equation}
see e.g. \cite{Ma}, Ex.~I.3.23. The role of moments is played by
the averages of images of the Schur functions under this map.
The representation theoretic definition of the spectral
$z$-measures implies that these averages can be explicitly
computed:
\begin{equation}\label{G}
\langle s_\lambda \rangle=(zz')^d\prod_{i=1}^d \frac{
(z+1)_{p_i}(z'+1)_{p_i}(-z+1)_{q_i}(-z'+1)_{q_i}}{p_i!q_i!}
\,{\det}\left(\frac{1}{p_i+q_j+1}\right)
\end{equation}
for any partition $\lambda=(p_1,\dots,p_d\mid q_1,\dots,q_d)$,
cf. (\ref{D}).

 Let us view pairs of sequences $(\alpha,\beta)$ as
point configurations
$$
(-\beta_1,-\beta_2,\dots,\alpha_2,\alpha_1)
$$
in $\R^*=\R\setminus\{0\}$. Then any spectral $z$-measure
defines a point process on $\R^*$. Formula (\ref{F}) defines a
map of $\Lambda$ to functions on such point configurations. It
is not hard to see that (\ref{G}) implies

\begin{claim4} The random point processes on $\R^*$ associated
with the spectral $z$-measures are Giambelli compatible.
\end{claim4}

This fact and the product formulas
$$
H(u)(\alpha,\beta)= \prod\limits_{i=1}^{\infty} \frac{1+\beta_i
u^{-1}}{1-\alpha_i u^{-1}}\,, \qquad E(v)(\alpha,\beta)=
\prod\limits_{i=1}^{\infty}\frac{1+\alpha_i v^{-1}}{1-\beta_i
v^{-1}}\,,
$$
which hold with probability 1, allow us to obtain the
determinantal formula for the correlation functions of the point
processes on $\R^*$ and to express the correlation kernel in
terms of the 2-point average $\langle H(u)E(v)\rangle$ by a
formula similar to (\ref{H}) with residues replaced by jumps
across the real axis. A straightforward computation leads to
explicit expressions for $\langle H(u)E(v)\rangle$ and, thus,
for the correlation kernel in terms of the confluent
hypergeometric functions or, equivalently, the Whittaker
functions.

This argument yields a relatively short derivation of the {\it
Whittaker kernel} obtained earlier by much heavier machinery in
\cite{B-AA}, see also \cite{BO-MRL}, \cite{BO-hyper}. Since the
Whittaker kernel essentially provides a complete solution to a
problem of harmonic analysis on the infinite symmetric group, we
see that the formalism of Giambelli compatible processes
delivers adequate tools for a direct solution of this problem.
Details and references on spectral $z$-measures can be found in
\S5.

\section{Preliminaries on Schur functions, multiparameter Schur
functions and Frobenius-Schur functions}

 In this section our main
references are Macdonald's book \cite{Ma} (symmetric functions
in general) and Olshanski, Regev and Vershik \cite{ORV}
(multiparameter Schur functions and Frobenius--Schur functions).

\subsection{Schur functions}
Let $\Lambda$ denote the algebra of symmetric functions. The
algebra $\Lambda$ can be considered as the algebra of
polynomials $\mathbb{C}[\bp_1,\bp_2\ldots ]$ in power sums
$\bp_1, \bp_2\ldots$. Then it can be realized, in different
ways, as an algebra of functions, depending on a specialization
of the generators $\bp_k$. The elements $h_k$ and $e_k$ (the
complete homogeneous symmetric functions and the elementary
symmetric functions) can be introduced through the generating
series:
$$
1+\sum\limits_{k=1}^{\infty}h_kt^k
=\exp\left(\sum\limits_{k=1}^{\infty}\bp_k\frac{t^k}{k}\right)=
\left(1+\sum\limits_{k=1}^{\infty}e_k(-t)^k\right)^{-1}.
$$
The Schur function $s_{\mu}$ indexed by a Young diagram $\mu$
can then be introduced through the Jacobi--Trudi formula:
$$
s_{\mu}=\det\left[h_{\mu_i-i+j}\right],
$$
where, by convention, $h_0=1, h_{-1}=h_{-2}=\ldots =0$, and the
order of the determinant is any number greater or equal to
$l(\mu)$ (the number of nonzero row lengths of $\mu$).

Define the generating series for $\{h_k\}$ and $\{e_k\}$ as formal
series in $\frac{1}{u}$ by
$$
H(u)=1+\sum\limits_{k=1}^{\infty}\frac{h_k}{u^k},\;\;\;E(u)=1+\sum\limits_{k=1}^{\infty}
\frac{e_k}{u^k}.
$$
For $p,q=0,1,\ldots $, let $(p|q)$ denote the hook Young diagram
$(p+1,1^q)$, and let $s_{(p|q)}$ be the ``hook'' Schur function
associated with this diagram. The following formula holds
(\cite{Ma}, Ex. I.3.14) :
\begin{equation}\label{11A}
H(u)E(v)=1+(u+v)\sum\limits_{p,q=0}^{\infty}\frac{s_{(p|q)}}{u^{p+1}v^{q+1}}.
\end{equation}
In the Frobenius notation, a Young diagram is written as
$$
\mu=\left(p_1,\ldots ,p_d |q_1,\ldots ,q_d\right),
$$
where $d$ is the number of diagonal boxes, and
$$
p_i=\mu_i-i,\;\;q_i=\mu_i'-i,
$$
where $\mu'$ is the transposed diagram.

In what follows we exploit the expression of the general Schur
functions through the hook Schur functions given by the
Giambelli formula (\cite{Ma}, Ex. I.3.9)
\begin{equation*}
s_{\mu}=\det\left[s_{(p_i|q_j)}\right]_{i,j=1}^d.
\end{equation*}
\subsection{Multiparameter Schur functions}
Let $a=(a_i)_{i\in\;\mathbb{Z}}$ be an arbitrary sequence of
complex numbers. The multiparameter analogs $h_{k,a}$ of the
complete homogeneous functions are introduced by the expression
\begin{equation*}
1+\sum\limits_{k=1}^{\infty}\frac{h_{k,a}}{(u-a_1)\ldots
(u-a_k)}=H(u)=1+\sum\limits_{k=1}^{\infty}\frac{h_k}{u^k}.
\end{equation*}
Since
$$
h_{k,a}=h_k+\mbox{lower terms},
$$
$\{h_{k;a}\}_{k=1,2,\ldots}$ is a system of algebraically
independent generators of $\Lambda$. We agree that
$$
h_{0;a}=1,\; h_{-1;a}=h_{-2;a}=\ldots =0.
$$
For $r\in\; \mathbb{Z}$, let $\tau^{r}\cdot a$ be the result of
shifting $a$ by $r$ digits to the left,
$$
\left(\tau^r a\right)_i=a_{i+r}.
$$
The multiparameter Schur function $s_{\mu;a}$ indexed by an
arbitrary Young diagram $\mu$ is defined by
$$
s_{\mu;a}=\det\left[h_{\mu_i-i+j;\tau^{1-j} a}\right],
$$
where the order of the determinant is any number greater or
equal to $l(\mu)$. From the above definition and from the result
of Macdonald \cite{Ma}, Example I.3.21, it is clear that the
multiparameter Schur functions $s_{\mu;a}$ satisfy the Giambelli
formula
\begin{equation}\label{12A}
s_{\mu;a}=\det\left[s_{(p_i|q_j);a}\right]_{i,j=1}^d,
\end{equation}
where the determinant has order $d=d(\mu)$ and $p_1,\dots ,p_d,
q_1,\dots,q_d$ denote the Frobenius coordinates of $\mu$.

As shown in \cite{ORV}, formula (\ref{11A}) can be generalized
as follows:
\begin{equation}\label{12B}
H(u)E(v)=1+(u+v)\sum\limits_{p,q=0}^{\infty}\frac{s_{(p|q);a}}{(u|a)^{p+1}(v|\hat{a})^{q+1}},
\end{equation}
where $\hat{a}$ stands for the ``dual'' sequence attached to
$a$, i. e.
$$
\hat{a}_i=-a_{-i+1},
$$
and
$$
(x|a)^m=
  \begin{cases}
    (x-a_1)\ldots (x-a_m), & m\geq 1, \\
    1 & m=0.
  \end{cases}
$$

\subsection{Frobenius--Schur functions}
The Frobenius--Schur functions are a special case of the
multiparameter Schur functions:
$$
Fs_{\mu}=s_{\mu ;a},\qquad a_i=i-1/2.
$$
By (\ref{12A}), the Frobenius-Schur functions satisfy the
Giambelli formula
\begin{equation*}
Fs_{\mu}=\det\left[Fs_{(p_i|q_j)}\right]_{i,j=1}^d,
\end{equation*}
and the next formula is a particular case of (\ref{12B}):
\begin{equation}\label{1A}
H(u)E(v)=1+(u+v)\sum\limits_{p,q=0}^{\infty}\frac{Fs_{(p|q)}}{(u-\frac12)\ldots
(u-\frac{2p+1}{2})(v-\frac12)\ldots (v-\frac{2q+1}{2}) }.
\end{equation}

\subsection{The Young graph} Let $\mathbb{Y}$ denote the set of all Young
diagrams including the empty diagram $\varnothing$. We regard
$\Y$ as the set of the vertices of a graph, called the {\em
Young graph\/} and denoted also by $\Y$. The edges of the graph
$\Y$ are couples of diagrams $(\mu,\la)$ such that $\la$ is
obtained from $\mu$ by adding a box (we denote this relation as
$\mu\nearrow\la$).

Let $\mbox{dim}(\mu,\lambda)$ be the number of all paths going
from a vertex $\mu$ to a vertex $\lambda$ with
$|\lambda|>|\mu|$. We agree that $\mbox{dim}(\mu,\mu)=1$, and
$\dim\lambda=\dim(\varnothing,\lambda)$. Clearly, if
$\mu\subset\la$ then $\dim(\mu,\la)$  is equal to the number of
the standard Young tableaux of skew shape $\lambda/\mu$, and if
$\mu$ is not contained in $\la$ then $\dim(\mu,\la)=0$.

\subsection{Polynomial functions on $\Y$} Given $\la\in\Y$, let
$a_1,\dots,a_d$, $b_1,\dots, b_d$ stand for its {\em modified
Frobenius coordinates\/}:
$$
a_i=\la_i-i+\tfrac12, \quad b_i=\la'_i-i+\tfrac12, \qquad
i=1,\dots,d.
$$
Following \cite{KO} (see also \cite{ORV}) we realize $\La$ as an
algebra of functions on $\Y$ using the following specialization
of the Newton power sums
$$
\bp_k(\la)=\sum\limits_{i=1}^da_i^k+(-1)^{k-1}\sum\limits_{i=1}^db_i^k,
$$
where  $\la$ ranges over $\Y$. Then each $f\in\La$ becomes a
function $f(\la)$ on $\Y$. Such functions were called in
\cite{KO} the {\em polynomial functions\/} on $\Y$.

In particular, the generating series for $\{h_k\}$ and $\{e_k\}$
take the form (see e.g. Macdonald \cite{Ma}, Ex.~1.3.23, and
Olshanski, Regev and Vershik \cite{ORV}):
$$
H(u)(\la)=\prod\limits_{i=1}^d\frac{u+b_i}{u-a_i}\, \qquad
E(v)(\la)=\prod\limits_{i=1}^d\frac{v+a_i}{v-b_i}\,.
$$
Note that these expressions are rational functions in $u$ or $v$
for any fixed $\la$.

The characterizing property of the Frobenius-Schur functions
that we are going to exploit is expressed by the relation
\begin{equation}\label{15A}
Fs_{\mu}(\la)=\frac{\dim(\mu,\la)\,n^{\downarrow m}}{\dim\la}\,,
\end{equation}
where $m=|\mu|$, $n=|\la|$, and
$$
n^{\downarrow m}=
  \begin{cases}
    n(n-1)\ldots (n-m+1), & n\geq m, \\
    0, & n<m.
  \end{cases}
$$

\section{Giambelli compatibility and point processes}

\begin{defn}
Assume $f\mapsto\langle f\rangle$ is a linear functional on the
algebra $\Lambda$ of symmetric functions, such that
$\langle1\rangle=1$. Let us say that $\langle\,\cdot\,\rangle$
is {\em Giambelli compatible\/}  if  for any Young diagram
$\lambda=(p_1,\ldots ,p_d|q_1,\ldots ,q_d)$
\begin{equation}\label{2A}
\left\langle s_{\lambda}\right\rangle=\det\left(\left\langle
s_{(p_i|q_j)}\right\rangle\right)_{i,j=1}^d.
\end{equation}
\end{defn}

\begin{prop} A linear functional $\langle\,\cdot\,\rangle$ is Giambelli
compatible in the sense of the above definition if and only if
for all $d=1,2,\dots$
\begin{equation}\label{2B}
\left\langle\det\left(\frac{H(u_i)E(v_j)}{u_i+v_j}\right)_{i,j=1}^d\right\rangle
=\det\left(\frac{\left\langle
H(u_i)E(v_j)\right\rangle}{u_i+v_j}\right)_{i,j=1}^d.
\end{equation}
\end{prop}

Here we regard $H(u_i)E(v_j)/(u_i+v_j)$ as elements of the
algebra
$$
\Lambda[[u_1^{-1},\dots,u_d^{-1},
v_1^{-1},\dots,v_d^{-1}]]_{loc},
$$
where the subscript ``loc'' means localization with respect to
$\prod(u_i^{-1}+v_j^{-1})$, which makes it possible to deal with
$$
\frac1{u_i+v_j}=\frac{u_i^{-1}v_j^{-1}}{u_i^{-1}+v_j^{-1}}.
$$

\begin{proof} Let us show that (\ref{2A}) implies (\ref{2B}).
Indeed, by (\ref{11A})
$$
\det\left(\frac{H(u_i)E(v_j)-1}{u_i+v_j}\right)_{i,j=1}^d
=\det\left(\sum\limits_{p_i,\;q_j=0}^{\infty}
\frac{s_{(p_i|q_j)}}{u_i^{p_i+1}v_j^{q_j+1}}\right)_{i,j=1}^d
$$
$$
\qquad\qquad\quad=\underset{q_1,\ldots
,\;q_d=0}{\sum\limits_{p_1,\ldots ,\;
p_d=0}^{\infty}}\frac{\det\left(s_{(p_i|q_j)}\right)_{i,j=1}^d}{u_1^{p_1+1}\ldots
u_d^{p_d+1}v_1^{q_1+1}\ldots v_d^{q_d+1}}.
$$
Applying $\langle\,\cdot\,\rangle$ to the both sides we obtain
$$
\left\langle\det\left(\frac{H(u_i)E(v_j)-1}{u_i+v_j}\right)_{i,j=1}^d\right\rangle=\underset{q_1,\ldots
,\;q_d=0}{\sum\limits_{p_1,\ldots ,\;
p_d=0}^{\infty}}\frac{\left\langle\det\left(s_{(p_i|q_j)}\right)_{i,j=1}^d\right\rangle}{u_1^{p_1+1}\ldots
u_d^{p_d+1}v_1^{q_1+1}\ldots v_d^{q_d+1}}
$$
$$
=\underset{q_1,\ldots ,\;q_d=0}{\sum\limits_{p_1,\ldots ,\;
p_d=0}^{\infty}}\frac{\det\left(\left\langle
s_{(p_i|q_j)}\right\rangle\right)_{i,j=1}^d}{u_1^{p_1+1}\ldots
u_d^{p_d+1}v_1^{q_1+1}\ldots
v_d^{q_d+1}}=\det\left(\sum\limits_{p_i,\;q_j=0}^{\infty}\frac{\left\langle
s_{(p_i|q_j)}\right\rangle}{u_i^{p_i+1}v_j^{q_j+1}}\right)_{i,j=1}^d
$$
$$
=\det\left(\left\langle\frac{H(u_i)E(v_j)-1}{u_i+v_j}\right\rangle\right)_{i,j=1}^d,\qquad\qquad\qquad\qquad\qquad
$$
where in the second equality we have used the Giambelli
compatibility assumption.

Now we aim to remove the $-1$'s. Let $A$ and $B$ denote the
$d\times d$ matrices with entries
$$
A(i,j)=\frac{H(u_i)E(v_j)-1}{u_i+v_j},\quad
B(i,j)=\frac1{u_i+v_j}, \qquad i,j=1,\dots,d.
$$
Next, let $A_{IJ}$ and $B_{IJ}$ denote their submatrices
corresponding to subsets $I, J\subset\{1,\dots,d\}$ with
$|I|=|J|$. The above argument shows that
$$
\left\langle\det A_{IJ}\right\rangle=\det\left\langle
A_{IJ}\right\rangle.
$$
Since $B$ has numerical entries, we have
$$
\left\langle\det(A+B)\right\rangle=\det\left\langle
A+B\right\rangle,
$$
as follows from the expansion
$$
\det(A+B)=\sum\limits_{I,J}\pm\det A_{IJ}\det B_{\bar I\bar J},
$$
where $\bar I$ stands for the complement to $I$ in
$\{1,\dots,d\}$.

Thus, we have proved that (\ref{2A}) implies (\ref{2B}).
Finally, the whole argument above can be inverted, which proves
the inverse implication.
\end{proof}

\begin{prop} The Giambelli
compatibility property (\ref{2A}) remains intact if we replace
in (\ref{2A}) the Schur functions $s_\lambda$ by the
multiparameter Schur functions $s_{\lambda;a}$. That is if we
require
\begin{equation}\label{2C}
 \langle s_{\lambda;a}\rangle=\det\left[\langle
s_{(p_i|q_j);a}\rangle\right]_{i,j=1}^d
\end{equation}
for any Young diagram $\lambda=(p_1,\ldots ,p_d\mid q_1,\ldots
,q_d)$.
\end{prop}

\begin{proof} Indeed, the transition formulas between multiparameter Schur
functions with different parameters (see \cite{ORV}, Theorem 7.3) imply that
conditions (\ref{2A}) and (\ref{2C}) are equivalent. Another way to see this is
to observe that the proof of Proposition 2.2 used only relations which hold
equally well for the multiparameter Schur functions.
\end{proof}

Assume $S$ is a Borel space equipped with a probability Borel
measure $P$. Let $\A(S,P)$ be the set of Borel functions $f$ on
$S$ such that $|f|, |f|^2, |f|^3,\dots$ belong to $L^1(S,P)$.
Clearly, $\A(S,P)$ is an algebra. Let
$\langle\,\cdot\,\rangle_P$ denote the expectation on $\A(S,P)$:
the linear functional determined by integration with respect to
measure $P$.

\begin{defn} Assume we are given an
algebra morphism $\phi: \Lambda\to\A(S,P)$. Let us say that the
triple $(S,P,\phi)$ is {\em Giambelli compatible\/} if the
pullback of $\langle\,\cdot\,\rangle_P$ on $\La$ is a Giambelli
compatible functional in the sense of Definition 2.1.
\end{defn}

Finally, recall some basic definitions related to random point
processes; for more detailed information, see Daley and
Vere--Jones \cite{DVJ} and Lenard \cite{Le}.

Let $\X$ be a locally compact space. By a {\em point
configuration\/} in $\X$ we mean a finite or countably infinite
collection of points of the space $\X$ with no accumulation
points. The set of all point configurations in $\X$ will be
denoted by $\CONF(\X)$; it admits a natural Borel structure. By
definition, a {\em random point process\/} on $\X$ is defined by
specifying a Borel map $S\to\CONF(\X)$, where $(S,P)$ is a Borel
space with a probability measure. Then the pushforward
$\mathcal{P}$ of $P$ is a probability measure on $\CONF(\X)$,
hence one can speak about {\em random\/}  point configurations
on $\X$. Since only the resulting measure $\mathcal{P}$ is
actually relevant, a point process is often viewed simply as a
couple $(\CONF(\X), \mathcal{P})$, the ``source'' probability
space $(S,P)$ being unnecessary or playing only an auxiliary
role. However, in the concrete examples we deal with in sections
4 and 5, the situation is somewhat different: we are primarily
interested in describing a measure $P$ on a space $S$ while the
point process generated by $(S,P)$ is  used rather as a tool.

The $m$th {\em correlation measure\/} $\rho_m$ ($m=1,2,\dots$)
of a random point process is a symmetric measure on
$\X^m=\X\times\dots\times\X$ ($m$ times) determined by
\begin{equation*}
\left\langle\sum\limits_{y_1,\dots,y_m\in X}F(y_1,\dots
,y_m)\right\rangle_{\mathcal{P}}=\int F(x_1,\dots
,x_m)\rho_m(dx_1\dots dx_m),
\end{equation*}
where the sum is taken over all ordered $m$-tuples of pairwise
distinct points taken from  the random point configuration $X$
and $F$ is a test function on $\X^m$.

The space $\X$ usually comes with a natural reference measure
$\nu(dx)$ such that $\rho_m$ is absolutely continuous with
respect to $\nu^{\otimes m}$ for all $m$. In such a case one can
consider the density of $\rho_m$ with respect to $\nu^{\otimes
m}$, which is called the $m$th {\em correlation function\/} of
the process. We will denote this function as
$\rho_m(x_1,\dots,x_m)$. The process is called {\em
determinantal\/} if there exists a function $K(x,y)$ on
$\X\times\X$ such that for any $m=1,2,\dots$
\begin{equation}\label{2D}
\rho_m(x_1,\dots,x_m)=\det\left(K(x_i,x_j)\right)_{i,j=1,\dots,m}\,.
\end{equation}

In our concrete examples, the point processes turn out to be
determinantal ones, and we will show how determinantal identity
(\ref{2B}), which holds for Giambelli compatible triples
$(S,P,\phi)$, leads to determinantal identity (\ref{2D}).

\section{The unitary ensemble of Random Matrix Theory}

\subsection{Basic notation}
Fix an arbitrary measure $\al$ on $\R$ with finite moments and
also fix $N=1,2,\dots$. In this section we take $\X=\R$ and
consider the subset $\CONF_N(\R)\subset\CONF(\R)$ consisting of
$N$--point configurations $X=(x_1,\dots,x_N)$. We also regard
$\CONF_N(\R)$ as the ``source'' space $S$. On this space we
define a probability measure $P_{\al,N}$, as follows:
$$
P_{\al,N}(dX)=\const_N\,V^2(x)\,\al^{\otimes N}(dX),
$$
where $\al^{\otimes N}(dX)=\prod_{i=1}^N\al(dx_i)$,
$V(X)=\prod_{1\leq i<j\leq N}(x_i-x_j)$ is the Vandermonde
determinant, and $\const_N$ is the normalization constant. For a
symmetric function $f(x_1,\ldots ,x_N)$ of the $x_i$'s, we
denote by $\langle f\rangle_{\al, N}$ its average with respect
to $P_{\al,N}$.

If we interpret the points $x_1, x_2,\ldots ,x_N$ of the random
point configuration $X$ as eigenvalues of a random $N\times N$
Hermitian matrix, then measure $P_{\al,N}$ determines a unitary
invariant ($\beta=2$) ensemble of Random Matrix Theory (see
Mehta \cite{Me}, Deift \cite{De} for details).

\subsection{Giambelli compatibility}
To any $f\in\La$ we assign a function $\phi(f)$ on
configurations $X$ in a natural way
$$
(\phi(f))(X)=f(x_1,\dots,x_N,0,0,\dots), \qquad
X=(x_1,\dots,x_N).
$$
Since, by assumption, all moments of $\al$ are finite,
$(\phi(f))(X)$ belongs to $\A(\CONF_N(\R),P_{\al,N})$. To
simplify the notation we will write $f(X)$ instead of
$(\phi(f))(X)$. Note that
$$
s_{\la}(X)=\frac{\det\left(x_i^{\lambda_j+N-j}\right)_{i,j=1}^N}
{V(x_1,\dots ,x_N)}, \quad \ell(\la)\le N; \qquad s_{\la}(X)=0,
\quad \ell(\la)>N,
$$
and
$$
H(u)(X)=\prod_{i=1}^N\frac{u}{u-x_i}\,, \qquad
E(v)(X)=\prod_{i=1}^N\frac{v+x_i}{v}.
$$

\begin{thm} The triple $(\CONF_N(\R),P_{\al,N},\phi)$ is
Giambelli compatible in the sense of Definitions 2.1 and 2.4.
\end{thm}

\begin{proof} Let $A_n$ denote the $n$th moment of $\al$,
$$
A_n=\int_{\R}x^n\,\al(dx), \qquad n=0,1,\dots.
$$
Assume first $\ell(\la)\le N$. The above expression for $s_\la(X)$ together
with the definition of $P_{\al,N}$ imply
\begin{multline*}
\langle s_\la\rangle_{\al,N}
=\const\,\int\dots\int\det(x_i^{\la_j+N-j})\det(x_i^{N-j})
\al(dx_1)\dots\al(dx_N)\\=\const \det(A_{\la_i+N-i+N-j}).
\end{multline*}
Here the second equality is obtained by a well--known trick, see, e.g., [58].
All determinants above are of order $N$.

Hence we obtain
$$
\langle s_\la\rangle_{\al,N}=
\begin{cases}\const\,\det(A_{\la_i+N-i+N-j})_{i,j=1}^N, & \ell(\la)\le
N,\\ 0, & \text{otherwise.} \end{cases}
$$

Now, our claim becomes a particular case of a general theorem due to Macdonald
(see \cite[Example I.3.21]{Ma}) which says:

Let $\{h_{rs}\}_{r\in\Z, \, s\in\Z_+}$ be any collection of commuting
indeterminates such that
$$
h_{0s}=1, \quad h_{-1,s}=h_{-2,s}=\dots=0 \qquad \forall s\in\Z_+,
$$
and set
$$
\wt s_\la=\det(h_{\la_i-i+j,\,j-1})_{i,j=1}^k
$$
where $k$ is any number $\ge\ell(\la)$. Then we have
$$
\wt s_\la=\det(\wt s_{(p_i\mid q_j)})_{i,j=1}^d,\qquad \la=(p_1,\dots,p_d\mid
q_1,\dots,q_d).
$$

To apply Macdonald's theorem consider the matrix $g=(g_{kl})$ of format
$\infty\times N$ with entries $g_{kl}=A_{k+l}$, where $k\in\Z_+$,
$l=0,\dots,N-1$. Next, multiply $g$ on the right by a suitable nondegenerate
matrix $N\times N$ in such a way that the resulting matrix $g'=(g'_{kl})$ be
strictly lower triangular:
$$
g'_{kl}=\delta_{kl}, \qquad 0\le k\le l\le N-1.
$$
Then
$$
\langle s_\la\rangle_{\al,N}=\det(g'_{\la_i+N-i,\,N-j})_{i,j=1}^N, \qquad
\ell(\la)\le N.
$$

Setting
$$
h_{rs}=\begin{cases} g'_{r-s+N-1,\,N-s-1}, & s=0,1,\dots,N-1, \, r\ge0\\
\de_{r0}, & s\ge N, \, r\ge0\\
0, & s\ge0, \, r<0,
\end{cases}
$$
it is readily seen that $\langle s_\la\rangle_{\al,N}$ coincides with
$\det(h_{\la_i-i+j,\,j-1})$ both for $\ell(\la)\le N$ and for $\ell(\la)>N$ (in
the latter case the determinant vanishes).
\end{proof}

\subsection{The correlation kernel}
It is well known (see, e.g., \cite{Me}) that the process
$(\CONF_N(\R),P_{\al,N})$ is determinantal and its correlation
kernel $K(x,y)$ is essentially the kernel of the projection
operator in $L^2(\R,\al)$ whose range is the space of
polynomials of degree $\le N-1$. The kernel can be written
explicitly in terms of orthogonal polynomials
$\pi_0,\pi_1,\dots$ corresponding to the weight $\al$.

Here we present a different expression for $K(x,y)$ which does
not involve orthogonal polynomials; instead of them we are
dealing with averages $\langle H(u)E(v)\rangle_{\al,N}$.

Assume for simplicity that $\al$ is a pure atomic measure. Then
we may speak about probability $\mbox{Prob}(X)$ of each
individual configuration $X$.  We may assume $X\subset\X$, where
$\X$ is a discrete subset of $\R$, the support of $\al$.

By definition, the $m$-point correlation function $\rho_m(Y)$,
where $Y=(y_1,\ldots ,y_m)$ is a subset of $\X$, is given by
\begin{equation}\label{32A}
\rho_m(Y)=\sum\limits_{X\supset Y}\mbox{Prob}(X).
\end{equation}
(To pass from correlation measures to correlation functions we
use the counting measure on $\X$ as the reference measure.)

\begin{prop} Let $\al$ be a pure atomic measure supported
by a discrete subset $\X\subset\R$. Then correlation functions
(\ref{32A}) are given by a determinantal formula
\begin{equation*}
\rho_m(Y)=\det(K(y_i,y_j))
\end{equation*}
with the correlation kernel
$$
K(x,y)=\underset{u=y}{\RES}\,\left(\frac{\langle
E(-x)H(u)\rangle_{\al,N}}{x-y}\right)
$$
(for $x=y$ the value of the kernel can be found using the L'Hospital rule).
\end{prop}

A proof of this result is given in \cite{BS}, \S2.8. To make a
connection with the notation of \cite{BS}, note that
$$
(E(-v)H(u))(X)=\frac{u^N}{v^N}\,\prod_{i=1}^N\frac{v-x_i}{u-x_i}\,.
$$
The argument of \cite{BS} relies on identity (\ref{2B}) (see formula 2.8.4 in
\cite{BS}). We can now obtain this identity as a direct corollary of Theorem
3.1 and Proposition 2.2.

As shown in \cite{BS}, \S2.8, the result of Proposition 3.2
implies the classical expression of the kernel in terms of
orthogonal polynomials.

A similar approach is presented in detail in the next section
for the more difficult case of $z$--measures.

Finally, it is worth noting that the assumption that $\al$ is
pure atomic can be removed. Then instead of residues of
functions with isolated singularities one has to deal with jumps
on a contour of functions which are holomorphic outside this
contour (in our case, the contour is the support of $\al$).

\section{$Z$-measures as Giambelli compatible point processes}

In this section, the ``source'' space $S$ is the set $\Y$ of
Young diagrams and as $P$ we take the so--called {\em
$z$-measures\/}. The related point processes live on the
discrete space $\Z'=\Z+\frac12$, the lattice of semi--integers.
We give only necessary definitions and refer to
(\cite{BO-hyper}, \cite{Ol-surv}) for motivation and details. We
show that the $z$--measures are Giambelli compatible. Using this
fact we then prove that the related lattice point processes are
determinantal, and we derive a formula for their correlation
kernel.

\subsection{The $z$--measures}
Let
$$
(a)_k=a(a+1)\ldots (a+k-1),\qquad (a)_0=1,
$$
denote the Pochhammer symbol. We fix two complex parameters $z$,
$z'$ such that the numbers $(z)_k(z')_k$ and $(-z)_k(-z')_k$ are
real and strictly positive for any $k=1,2,\ldots$. These
assumptions on $z,z'$ are satisfied  if and only if one of the
following two conditions holds:
\begin{itemize}
    \item either $z'=\bar{z}$ and $z\in \mathbb{C}\setminus \mathbb{Z}$
    \item or $z$, $z'\in \mathbb{R}$ and there exists $m\in
    \mathbb{Z}$ such that $m<z,z'<m+1$.
\end{itemize}

Let $\Y_n$ denote the finite set of Young diagrams with $n$
boxes $(n=1,2\dots)$. The {\em $z$--measure\/} on $\Y_n$ with
parameters $z,z'$ is defined by
\begin{equation}\label{41A}
M_{z,z'}^{(n)}(\lambda)=
\frac{\prod\limits_{(i,j)\in\lambda}(z+j-i)(z'+j-i)}{(zz')_n}
\;\frac{\left(\dim\lambda\right)^2}{n!},\qquad \lambda\in
\mathbb{Y}_n,
\end{equation}
where ``$(i,j)\in\la$'' stands for the box of diagram $\la$ with
row coordinate $i$ and column coordinate $j$, and $\dim\lambda$
denotes the number of the standard Young tableaux of shape
$\lambda$. This is a probability measure.

Further, let $\xi\in(0,1)$ be an additional parameter. The {\em
mixed $z$--measure\/} with parameters $z,z',\xi$ is a
probability measure on the set $\Y$ of all Young diagrams
defined by
$$
M_{z,z',\xi}(\lambda)=
M_{z,z'}^{(n)}(\lambda)\cdot(1-\xi)^{zz'}\frac{\left(zz'\right)_n}{n!}\,
\xi^n,\qquad n=|\la|,
$$
where $M_{z,z'}^{(0)}(\varnothing):=1$.

\subsection{Giambelli compatibility} Recall
that $\La$ can be viewed as the algebra of ``polynomial
functions'' on $\Y$.  We write $\langle f\rangle_{M_{z,z',\xi}}$
for the expectation of a function $f$ with respect to the
probability measure $M_{z,z',\xi}$. It turns out that the
quantities $\langle f\rangle_{M_{z,z',\xi}}$ are readily
computed for the Frobenius--Schur functions $Fs_\mu$.

\begin{prop}
For any $\mu\in \mathbb{Y}$,
\begin{equation}\label{42B}
\left\langle
Fs_{\mu}\right\rangle_{M_{z,z',\xi}}=\left(\frac{\xi}{1-\xi}\right)^{|\mu|}
\prod\limits_{(i,j)\in\mu}(z+j-i)(z'+j-i)\cdot\frac{\dim\mu}{|\mu|!}.
\end{equation}
\end{prop}

\begin{proof} The computation relies on formula (\ref{15A}).
First of all, note that this formula implies that
$Fs_\mu(\la)\ge0$ for any $\la$, which justifies transformations
of infinite sums below.

{}From (\ref{15A}) we obtain
\begin{equation}\label{42A}
\left\langle
Fs_{\mu}\right\rangle_{M_{z,z',\xi}}=\left(1-\xi\right)^{zz'}\sum\limits_{n=m}^{\infty}
\frac{(zz')_n\xi^n\,n^{\downarrow m}}{n!}\sum\limits_{\lambda\in
\mathbb{Y}_{n}}\dim(\mu,\lambda)\frac{M^{(n)}_{z,z'}(\la)}{\dim\la}\,,
\end{equation}
where $m=|\mu|$. Now we use the fact that the function
$$
\varphi_{z,z'}(\la):=\frac{M^{(|\la|)}_{z,z'}(\la)}{\dim\la}
$$
is {\em harmonic\/} on the Young graph in the sense of Vershik
and Kerov \cite{VK}. That is,
$$
\varphi_{z,z'}(\mu)=\sum_{\la:\,
\mu\nearrow\la}\varphi_{z,z'}(\la)\qquad \forall\mu\in\Y_m,
$$
see \cite{PartI}, \cite{KOV}, \cite{Ke}, \cite{Ok-SL2} for different proofs.
Iterating this relation we obtain
$$
\varphi_{z,z'}(\mu)=\sum_{\la\in\Y_n}\dim(\mu,\la)\varphi_{z,z'}(\la)\qquad
\forall n\ge m.
$$
Plugging this into (\ref{42A}) gives
\begin{multline*}
\left\langle
Fs_{\mu}\right\rangle_{M_{z,z',\xi}}=\frac{M^{(|\mu|)}_{z,z'}(\mu)}{\dim\mu}\,
(1-\xi)^{zz'} \sum\limits_{n\geq
m}(zz')_n\xi^n\frac{n^{\downarrow m}}{n!}\\
=\frac{M^{(m)}_{z,z'}(\mu)}{\dim\mu}\, (1-\xi)^{zz'}
(zz')_m\xi^m\sum\limits_{n\geq
m}\frac{(zz'+m)_{n-m}\,\xi^{n-m}}{(n-m)!}\,.
\end{multline*}
Finally, observe that the latter sum equals $(1-\xi)^{-zz'-m}$,
and use the explicit expression for $M^{(m)}_{z,z'}(\mu)$ (see
(\ref{41A})). This gives (\ref{42B}).
\end{proof}

The first consequence of Proposition 4.1 is that all functions
$f(\la)$, where $f\in\La$, are summable with respect to
$M_{z,z',\xi}$. Therefore, the map $f\mapsto f(\,\cdot\,)$
determines a morphism $\phi:\La\to\A(\Y, M_{z,z',\xi})$.

\begin{prop} The triple
$(\Y, M_{z,z',\xi}, \phi)$ is Giambelli compatible.
\end{prop}

\begin{proof}
Let us show that
\begin{equation}\label{42C}
\left\langle Fs_{\mu}\right\rangle_{M_{z,z',\xi}}=
\det\left[\left\langle
Fs_{(p_i|q_j)}\right\rangle_{M_{z,z',\xi}}\right]_{i,j=1}^d
\end{equation}
Indeed, we can rewrite (\ref{42B}) in terms of Frobenius
coordinates: for the product over the boxes this is easy, and
for the dimension we use the formula
$$
\frac{\dim\mu}{|\mu|!}=\frac{\prod\limits_{i<j}(p_i-p_j)(q_i-q_j)}
{\prod\limits_{i}p_i!q_i!\cdot\prod\limits_{i,j}(p_i+q_j+1)}
=\det\left[\frac1{p_i!q_j!(p_i+q_j+1)}\right]
$$
(see, e.g., \cite{PartI}). Then we obtain (\ref{42C}) which in
turn implies the claim, by virtue of Proposition 2.3.
\end{proof}

\subsection{Computation of $\left\langle
H(u)E(v)\right\rangle_{M_{z,z',\xi}}$} Below  $u$ and $v$ are
assumed to be complex variables (rather than formal parameters,
as in \S2). To ensure the existence $\left\langle
H(u)E(v)\right\rangle_{M_{z,z',\xi}}$ we need an estimate of
$(H(u)E(v))(\la)$. It is provided by the next lemma, which is
stated in a slightly greater generality, because we will need a
similar estimate in \S5.

\begin{lem}
Assume that $u$ is a complex variable subject to constraints
$$
\eps<|\arg(u)|<\pi-\eps
$$
with a certain $\eps>0$. Take $\de>0$ and two infinite sequences
$\al_1\ge\al_2\ge\dots\ge0$, $\be_1\ge\be_2\ge\dots\ge0$ such
that
$$
\sum_{i=1}^\infty(\al_i+\be_i)\le\de.
$$
Then
$$
\left|\prod_{i=1}^\infty\frac{1+\be_iu^{-1}}{1-\al_iu^{-1}}\right|
\le e^{C\,\de\,|u^{-1}|},
$$
where $C=C(\eps)>0$ is a constant depending only on $\eps$.
\end{lem}

\begin{proof}
The numerator admits a trivial estimate,
$$
|1+\be_i u^{-1}|\le 1+|\be_i u^{-1}|\le e^{\be_i\,|u^{-1}|},
$$
which implies
$$
\left|\prod_{i=1}^\infty (1+\be_iu^{-1})\right|\le
e^{\de\,|u^{-1}|}.
$$

As for the denominator, we will estimate it separately for $i\in
I$ and for $i\notin I$, where
$$
I=\{i\,\mid\, \al_i|u^{-1}|\ge \tfrac12\}.
$$

Assume first $i\notin I$. Then $\al_i|u^{-1}|\le\tfrac12$,
whence
$$
\left|\frac1{1-\al_i u^{-1}}\right|\le \frac1{1-\al_i|u^{-1}|} \le
1+2\al_i|u^{-1}|,
$$
because
$$
\frac1{1-x}\le 1+2x, \qquad 0\le x\le\tfrac12.
$$
Therefore,
$$
\left|\prod_{i\notin I}\frac1{1-\al_i u^{-1}}\right| \le \prod_{i\notin
I}(1+2\al_i|u^{-1}|)\le e^{2\de\,|u^{-1}|}.
$$
Now assume that $i\in I$. Then $\al_i|u^{-1}|\ge\frac12$ and
$\al_i\ge\frac12|u|$. Since the sum of all $\al_i$ does not
exceed $\de$, we obtain that $|I|$ (the cardinality of $I$) does
not exceed $2\de|u^{-1}|$. Next, the constraints on $\arg(u)$
imply that $|\im u^{-1}|\ge C_1|u^{-1}|$ with a certain constant
$C_1$ depending only on $\eps$. Therefore
$$
\al_i|\im u^{-1}|\ge \al_i\, C_1\,|u^{-1}|,
$$
whence
$$
\left|\frac1{1-\al_i u^{-1}}\right|\le \frac1{\al_i|\im
u^{-1}|}\le \frac2{C_1}
$$
and
$$
\left|\prod_{i\in I}\frac1{1-\al_i u^{-1}}\right|\le
\left(\frac2{C_1}\right)^{|I|}\le
\left(\frac2{C_1}\right)^{2\de\,|u^{-1}|} \le
e^{C_2\,\de\,|u^{-1}|}.
$$

Combining all these estimates we obtain the desired inequality
with $C=3+C_2$.
\end{proof}

\begin{cor}
Fix $\eps>0$ and let $u_1,\dots,u_m,v_1,\dots,v_m$ be complex
variables subject to constraints
$$
\eps<|\arg(u_i)|<\pi-\eps, \quad \eps<|\arg(v_i)|<\pi-\eps,
\qquad i=1,\dots, m,
$$
and such that $|u_i|$, $|v_i|$ are large enough (greater than a
constant depending on $\eps$ and $\xi$). Then
$$
\sum_{\la\in\Y}|(H(u_1)E(-v_1)\dots
H(u_m)E(-v_m))(\la)|\,M_{z,z',\xi}(\la)<\infty.
$$
\end{cor}

\begin{proof}
Recall that
$$
H(u)(\la)=\prod_{j=1}^d\frac{1+b_j u^{-1}}{1-a_j u^{-1}}, \qquad
E(-v)(\la)=\prod_{j=1}^d\frac{1-a_j v^{-1}}{1+b_j v^{-1}}
$$
and note that $\sum_j(a_j+b_j)=n:=|\la|$. By virtue of Lemma
4.3,
$$
|(H(u_1)E(-v_1)\dots H(u_m)E(-v_m))(\la)|\le e^{Cn\sum_i
(|u_i^{-1}|+|v_i^{-1}|)}
$$
with the same constant $C=C(\eps)$ as in the lemma. Then it
follows from the definition of $M_{z,z',\xi}$  that the sum in
question is finite provided that $u_i$ and $v_i$ are so large
that
$$
e^{C\sum_i (|u_i^{-1}|+|v_i^{-1}|)} \xi<1.
$$
\end{proof}

Below we use the standard notation $F(a,b;c;\zeta)$ for the
Gauss hypergeometric function with parameters $a,b,c$ and
argument $\zeta$. Recall that $F(a,b;c;\zeta)$ is well defined
for $\zeta\in\C\setminus[1,+\infty)$. Moreover,
$F(a,b;c;\zeta)/\Ga(c)$ is an entire function of parameters
$(a,b,c)\in\C^3$. In particular, $F(a,b;c;\zeta)$ is a
meromorphic function in $c$ with poles at $c=0,-1,-2,\dots$.

By Lemma 4.3, the average $\left\langle
H(u)E(v)\right\rangle_{M_{z,z',\xi}}$ is well defined and is an
analytic function in $(u,v)$, provided that $(u,v)$ range over a
suitable domain in $\C^2$.

\begin{prop}
The average $\left\langle H(u)E(v)\right\rangle_{M_{z,z',\xi}}$
is given by the formula:
\begin{multline}\label{43A}
\left\langle H(u)E(v)\right\rangle_{M_{z,z',\xi}}=
F(z,z';-u+\tfrac12;\tfrac{\xi}{\xi-1})
F(-z,-z';-v+\tfrac12;\tfrac{\xi}{\xi-1})\\
+\frac{zz'\xi}{(1-\xi)^2(u-\tfrac12)(v-\tfrac12)}\\
\times F(z+1,z'+1;-u+\tfrac32;\tfrac{\xi}{\xi-1})
F(-z+1,-z'+1,-v+\tfrac32;\tfrac{\xi}{\xi-1}).
\end{multline}
\end{prop}

\begin{proof} By (\ref{12B}),
\begin{equation}\label{43B}
(H(u)E(v))(\la)=1+(u+v)\sum\limits_{p,\;q\geq 0}\frac{Fs_{(p\mid
q)}(\la)}{(u-\tfrac12)^{\downarrow p+1}(v-\tfrac12)^{\downarrow
q+1}}.
\end{equation}
Since we know $\langle Fs_{(p|q)}\rangle_{M_{z,z',\xi}}$, it is
tempting to average this relation over $\la$'s. However, we have
to be careful at this point, because now we are dealing with
actual functions in $(u,v)$  (not with formal series in $u^{-1},
v^{-1}$, as in \S2). Moreover, we cannot even expect that the
resulting expression would possess an expansion at
$(u,v)=(\infty,\infty)$, because, for a fixed $\xi$,  the
right--hand side of (\ref{43A}) is not a meromorphic function
near $(u,v)=(\infty,\infty)$ and hence does not admit such an
expansion.

This difficulty can be overcome using the following trick: we
will regard $\xi$ not as a numeric parameter but as a formal
indeterminate. Observe that both sides of (\ref{43A}) are
analytic functions in $\xi$ near $\xi=0$ such that the
coefficients of the Taylor expansion at $\xi=0$ are rational
functions in $(u,v)$ admitting an expansion at
$(u,v)=(\infty,\infty)$ (in more detail, these rational
functions are finite sums $\sum f_i(u)g_i(v)$, where $f_i$ and
$g_i$ are rational functions in one variable). Thus, we may
prove (\ref{43A}) as an identity in the algebra of formal power
series in $u^{-1}$, $v^{-1}$, and $\xi$. This provides a
justification for the formal computation below.

By (\ref{42B}), for the hook diagram $\mu=(p+1,1^{q})$ we have
\begin{equation*}
\left\langle
Fs_{(p|q)}\right\rangle_{M_{z,z',\xi}}=\left(\frac{\xi}{\xi-1}\right)^{p+q+1}
\;\frac{zz'\cdot (z+1)_p(z'+1)_p(-z+1)_q(-z'+1)_q}{p!q!(p+q+1)}.
\end{equation*}
Denote the expression in the right--hand side  by $A(p,q)$. Then
we obtain from (\ref{43B})
$$
\left\langle
H(u)E(v)\right\rangle_{M_{z,z',\xi}}=1+(u+v)\sum\limits_{p,\;q\geq
0}\frac{A(p,q)}{(u-\tfrac12)^{\downarrow
p+1}(v-\tfrac12)^{\downarrow q+1}}.
$$
Write
$$
u+v=(u-p-\tfrac12)+(v-q-\tfrac12)+(p+q+1),
$$
and plug this expression into the sum. Then we obtain
\begin{equation*}
\left\langle
H(u)E(v)\right\rangle_{M_{z,z',\xi}}=1+\sum\limits_{p,\;q\geq
0}\frac{A(p,q)}{(u-\tfrac12)^{\downarrow
p}(v-\tfrac12)^{\downarrow q+1}} +\sum\limits_{p,\;q\geq
0}\frac{A(p,q)}{(u-\tfrac12)^{\downarrow
p+1}(v-\tfrac12)^{\downarrow q}}
\end{equation*}
$$
+\sum\limits_{p,\;q\geq
0}\frac{(p+q+1)A(p,q)}{(u-\tfrac12)^{\downarrow
p+1}(v-\tfrac12)^{\downarrow q+1}}.
$$
Decompose the first sum into two parts in such a way that the
first part corresponds to summation over index $q$ with $p$
being equal to zero, while the second part is summation over
$p\geq 1$ and $q\geq 0$. Replace index $p$ by $p-1$ in the
second part, then $A(p,q)$ is replaced by $A(p+1,q)$. Decompose
the second sum  in the same way, and obtain
\begin{equation*}
\left\langle
H(u)E(v)\right\rangle_{M_{z,z',\xi}}=1+\sum\limits_{q\geq
0}\frac{A(0,q)}{(v-\tfrac12)^{\downarrow q+1}}
+\sum\limits_{p\geq 0}\frac{A(p,0)}{(u-\tfrac12)^{\downarrow
p+1}}
\end{equation*}
$$
+\sum\limits_{p,\;q\geq
0}\frac{(p+q+1)A(p,q)+A(p,q+1)+A(p+1,q)}{(u-\tfrac12)^{\downarrow
p+1}(v-\tfrac12)^{\downarrow q+1}}.
$$
The first two sums can be immediately rewritten in terms of
hypergeometric functions. To compute the last sum we observe that
\begin{multline*}
A(p+1,q)+A(p,q+1)+(p+q+1)A(p,q)\\=
\left(\frac{\xi}{1-\xi}\right)^{p+q+2}
\frac{(z)_{p+1}(z')_{p+1}(-z)_{q+1}(-z')_{q+1}}{(p+1)!(q+1)!}\\
+\frac1{1-\xi}\left(\frac{\xi}{1-\xi}\right)^{p+q+1}
\frac{zz'(z+1)_{p}(z'+1)_{p}(-z+1)_{q}(-z'+1)_{q}}{p!q!}\,.
\end{multline*}
It follows that the last sum can be decomposed into two sums and
rewritten in terms of hypergeometric functions. With these
preparations we find
\begin{gather*}
\left\langle H(u)E(v)\right\rangle_{M_{z,z',\xi}}=1+
\left(F(-z,-z';-v+\tfrac12;\tfrac{\xi}{1-\xi})-1\right)\\+
\left(F(z,z';-u+\tfrac12;\tfrac{\xi}{1-\xi})-1\right)
\\
+\left(F(-z,-z';-v+\tfrac12;\tfrac{\xi}{1-\xi})-1\right)
\left(F(z,z';-u+\tfrac12;\tfrac{\xi}{1-\xi})-1\right)
\\
+\frac{\xi\,zz'}{(1-\xi)^2(u-\tfrac12)(v-\tfrac12)}
F(z+1,z'+1;-u+\tfrac32;\tfrac{\xi}{\xi-1})\\
\times F(-z+1,-z'+1;-v+\tfrac32;\tfrac{\xi}{\xi-1}).
\end{gather*}
After simplifications we obtain desired formula (\ref{43A}).
\end{proof}

\subsection{Correlation measures and controlling measures} Set
$$
\mathbb{Z}'=\mathbb{Z}+\frac{1}{2}
=\left\{\ldots,-\frac{3}{2},-\frac{1}{2},\frac{1}{2},\frac{3}{2},\ldots\right\}.
$$
As in \S1.5, for any $\lambda\in \mathbb{Y}$ we define the
modified Frobenius coordinates of $\lambda$ as
$$
a_i=a_i(\la)=p_i+\tfrac{1}{2}=\lambda_i-i+\tfrac{1}{2},\quad
b_i=b_i(\la)=q_i+\tfrac{1}{2}=\lambda'_i-i+\tfrac{1}{2},
$$
where $i=1,\dots,d$ and $d=d(\la)$ denotes the number of
diagonal boxes in $\lambda$. Note that
$$
a_1>\ldots >a_d>0,\quad b_1>b_2>\ldots
>b_d>0,\quad\sum\limits_{i=1}^d(a_i+b_i)=|\lambda|.
$$
Using this notation we assign to an arbitrary Young diagram a
point configuration $X=X(\la)\in\;\CONF(\Z')$, as follows
$$
X(\la)=\left\{-b_1,\ldots,-b_d,a_d,\ldots,a_1 \right\}.
$$

Note that a point configuration $X$ on $\Z'$ comes from a Young
diagram $\la$ if and only if $X$ is finite and {\em balanced\/}
in the sense that it has equally many points to the left and to
the right of zero.

Thus, the correspondence $\lambda\mapsto X(\la)$ defines a
bijection between Young diagrams $\la$ and balanced
configurations $X$, and we will often identify $\la$ and
$X(\la)$. Assume we are given a probability measure $M$ on $\Y$.
Then  we obtain a point process on $\Z'$ with ``source space''
$(\Y,M)$. Let $\rho_m$ stand for the $m$th correlation measure
of this process; $\rho_m$ is supported by the subset
$$
(\Z')^m_0=\{(x_1,\dots,x_m)\in(\Z')^m\mid x_i\ne x_j, \quad i\ne
j\}.
$$

Since $\Z'$ is a discrete space, there is no essential
difference between correlation measures and correlation
functions (see the end of section 2; here we take the counting
measure on $\Z'$ as the reference measure). Note that
$\rho_m(x_1,\dots,x_m)$ is the probability that the random
configuration contains $\{x_1,\dots,x_m\}$.

We will introduce one more concept, that of {\em controlling
measures\/} \cite{PartI}. The definition is as follows. First,
to an arbitrary $\la\in\Y$ we assign a measure on $\Z'$:
$$
\si_\la=\sum_{i=1}^d (a_i\de_{a_i} + b_i\de_{-b_i}),
$$
where $a_i, b_i$ are the modified Frobenius coordinates of $\la$
and $\de_x$ stands for the delta measure at $x\in\Z'$. Second,
for any $m=1,2,\dots$ we take the $m$th power
$(\si_\la)^{\otimes m}$, which is a measure on $(\Z')^m$, and
then average it with respect to our initial probability measure
$M$:
$$
\si_m=\left\langle (\si_{\cdot})^{\otimes
m}\right\rangle_M=\sum_{\la\in\Y}(\si_\la)^{\otimes m}M(\la).
$$

\begin{lem} We have
$$
\si_m=|x_1\dots x_m|(\rho_m+\dots),
$$
where the dots denote a measure supported by
$(\Z')^m\setminus(\Z')^m_0$. In particular, on $(\Z')^m_0$, the
measure $\si_m$ coincides with the measure $\rho_m$ multiplied
by the function $|x_1\dots x_m|$.
\end{lem}

\begin{proof} Assume first that $M$ is the delta measure at a
point $\la\in\Y$. Then we have
\begin{multline*}
\si_m=\sum_{x_1,\dots,x_m\in X(\la)}|x_1\dots
x_m|\,\de_{x_1}\otimes\dots\otimes\de_{x_m}\,, \\
\rho_m=\sum_{\substack{x_1,\dots,x_m\in X(\la)\\\text{pairwise
distinct}}}\de_{x_1}\otimes\dots\otimes\de_{x_m}\,.
\end{multline*}
Clearly, this implies the desired equality in the special case
$M=\de_\la$. In the general case, both $\si_m$ and $\rho_m$ are
obtained from these expressions by averaging with respect to
$M$, which completes the proof.

A detailed description of the ``rest measure'' supported by
$(\Z')^m\setminus(\Z')^m_0$ is given in \cite{PartI}.
\end{proof}

Recall that the {\em Cauchy transform\/} of a measure $\nu$ on
$\R^m$ is given by
$$
\widehat\nu(u_1,\dots,u_m)=\int_{\R^m}\frac{\nu(dx)}{(u_1-x_1)\dots(u_m-x_m)}\,,
\qquad (u_1,\dots,u_m)\in(\C\setminus\R)^m.
$$
It is well defined if $\nu$ satisfies the {\em growth
condition\/}
\begin{equation}\label{44A}
\int_{\R^m}\frac{\nu(dx)}{(1+|x_1|)\dots(1+|x_m|)}<\infty.
\end{equation}
Note that the initial measure $\nu$ can be reconstructed from
its transform $\widehat\nu$.

In particular, if $\nu$ is a pure atomic measure whose support
has no accumulation points, then $\widehat\nu$ is a meromorphic
function in each variable $u_i$, and for any point
$(x_1,\dots,x_m)$ in the support of $\nu$, we have
\begin{equation}\label{44B}
\nu(x_1,\dots,x_m)=\underset{u_1=x_1}{\RES}\dots
\underset{u_m=x_m}{\RES} \widehat\nu(u_1,\dots,u_m).
\end{equation}

\begin{lem} Let $M$ be a probability measure on $\Y$ and $\si_m$
be the corresponding $m$--th controlling measure on $(\Z')^m$.
Assume that $\si_m$ satisfies the growth condition ensuring the
existence of the  Cauchy transform
$\widehat\si_m(u_1,\dots,u_m)$. Further, assume that the average
$$
\langle H(u_1)E(-v_1)\dots H(u_m)E(-v_m)\rangle_M
$$
is well defined for $(u_1,\dots,u_m)$ and $(v_1,\dots,v_m)$ ranging over a
domain $\mathcal D\subset(\C^\setminus\R)^m$.

Then for any $(u_1,\dots,u_m)\in\mathcal D$ we have
\begin{multline*}
\widehat\si_m(u_1,\dots,u_m)=u_1\dots u_m\\
\times\left\{\frac{\partial^m}{\partial v_1\dots\partial
v_m}\langle H(u_1)E(-v_1)\dots
H(u_m)E(-v_m)\rangle_M\right\}_{\substack{v_1=u_1\\
\dots\\ v_m=u_m}}
\end{multline*}
\end{lem}

\begin{proof}
Assume first that $M$ is the delta measure at $\la\in\Y$. Then
$$
\si_m=\si_\la^{\otimes m}=\bigg(\sum_{x\in
X(\la)}|x|\,\de_x\bigg)^{\otimes m},
$$
whence
$$
\widehat\si_m(u_1,\dots,u_m)=\sum_{x_1,\dots,x_m\in X(\la)}
\frac{|x_1\dots x_m|}{(u_1-x_1)\dots(u_m-x_m)}\,.
$$

On the other hand,
\begin{multline*}
\langle H(u_1)E(-v_1)\dots H(u_m)E(-v_m)\rangle_M\\
=(H(u_1)E(-v_1)\dots H(u_m)E(-v_m))(\la)\\
=\prod_{i=1}^m\prod_{j=1}^d \frac{(1+b_j u_i^{-1})(1-a_j
v_i^{-1})}{(1-a_j u_i^{-1})(1+b_j v_i^{-1})}\,.
\end{multline*}
Differentiating over $v_1,\dots,v_m$ and then specializing
$v_i=u_i$ for all $i=1,\dots,m$ gives
$$
\prod_{i=1}^m\left(\frac1{u_i}\,
\sum_{j=1}^d\left(\frac{a_j}{u_i-a_j}+\frac{b_j}{u_i+b_j}\right)\right),
$$
which leads to the same result after multiplication by $u_1\dots
u_m$.

Thus, we have verified the desired relation for $M=\de_\la$. In
the general case, we average over $\la$'s with respect to
measure $M$. To justify the interchange of the operation
``differentiation over $v_i$'s followed by specialization
$v_i=u_i$'' with the averaging operation, we observe that the
former operation can be written as a multiple contour
Cauchy--type integral.
\end{proof}

Let us abbreviate
$$
F_m(u_1,\dots,u_m)=\left\{\frac{\partial^m}{\partial
v_1\dots\partial v_m}\langle H(u_1)E(-v_1)\dots
H(u_m)E(-v_m)\rangle_M\right\}_{\substack{v_1=u_1\\
\dots\\ v_m=u_m}}
$$
By the above lemma, this function, which is initially defined in
a domain $\mathcal D\subset(\C\setminus\R)^m$, can be extended
to the whole $\C^m$ as a function which is meromorphic in each
variable $u_i$ with possible poles at points of the lattice
$\Z'$.

\begin{cor}
Under the hypotheses of the Lemma 4.7, for any $m$--tuple of
pairwise distinct numbers $x_1,\dots,x_m\in\Z'$,
$$
\rho_m(x_1,\dots,x_m)=\sgn(x_1)\dots\sgn(x_m)\,\underset{u_1=x_1}{\RES}\dots
\underset{u_m=x_m}{\RES} F_m(u_1,\dots,u_m),
$$
where $\sgn(x)=1$ for $x>0$ and $\sgn(x)=-1$ for $x<0$.
\end{cor}

\begin{proof} Indeed, by (\ref{44B}),
$$
\si_m(x_1,\dots,x_m)=\underset{u_1=x_1}{\RES}\dots
\underset{u_m=x_m}{\RES} \widehat\si_m(u_1,\dots,u_m).
$$
Then we apply Lemma 4.7.
\end{proof}

\subsection{Computation of the correlation functions}
Here we apply the above result to computing  the correlation
functions for the point process determined by $M=M_{z,z',\xi}$.

First of all, it should be noted that the two assumptions on $M$
made in Lemma 4.7 are satisfied for $M=M_{z,z',\xi}$.

In more detail, one of the assumptions was the growth condition
on $\si_m$. We claim that for $M=M_{z,z',\xi}$, the measure
$\si_m$ actually satisfies a stronger condition: it is a finite
measure. To see this, we observe that $\si_\la$ has mass
$\sum_i(a_i+b_i)=|\la|$, hence $\si_\la^{\otimes m}$ has mass
$|\la|^m$. Averaging over $\la$'s and recalling the definition
of $M_{z,z',\xi}$ we obtain that the total mass of $\si_m$
equals
$$
(1-\xi)^{zz'}\,\sum_{n=0}^\infty
n^m\,\frac{(zz')_n\,\xi^n}{n!}<\infty.
$$

Another assumption was that the average of
$$
(H(u_1)E(-v_1)\dots H(u_m)E(-v_m))(\la)
$$
with respect to $M$ exists provided that
$(u_1,\dots,u_m)$ and $(v_1,\dots,v_m)$ range in a suitable
domain $\mathcal D$. For $M=M_{z,z',\xi}$ this is indeed true
due to the estimate established in Lemma 4.3. As $\mathcal D$
one can take any domain of the form
$$
\eps<|\arg(u_i)|<\pi-\eps, \quad |u_i|\gg0, \qquad i=1,\dots,m.
$$

\begin{thm} The point process on $\Z'$ corresponding to the measure $M_{z,z',\xi}$ is
determinantal and its correlation kernel can be  written as
\begin{equation}\label{45C}
K(x,y)=\left\{
\begin{array}{ll}
   \underset{u=y}{\RES}\;\dfrac{\langle E(-x)H(u)\rangle}{x-y}, & x>0, y>0,\; x\neq y, \\
   & \\
    -\underset{u'=x}{\RES}\;\underset{u=y}{\RES}\;
    \dfrac{\langle E(-u')H(u)\rangle}{x-y}, &  x<0, y>0,\\
    & \\
    \dfrac{\langle E(-x)H(y)\rangle}{x-y}, & x>0, y<0, \\
    & \\
   -\underset{u=x}{\RES}\;\dfrac{\langle E(-u)H(y)\rangle}{x-y}, & x<0, y<0,\; x\neq y,  \\
\end{array}
\right.
\end{equation}
where $\langle\,\cdot\,\rangle$ means
$\langle\,\cdot\,\rangle_{M_{z,z',\xi}}$, and the indeterminacy
arising for $x=y$ is resolved via the L'Hospital rule.
\end{thm}

The statement of the theorem needs a few comments:

1) By Proposition 4.5, the quantity $\langle E(-v)H(u)\rangle$,
which is initially defined (as a function in $(u,v)$) in a
domain of $\C^2$, actually can be extended to a meromorphic
function on the whole $\C^2$. In the above formula for the
kernel we use this meromorphic extension.

2) Note that $\langle E(-v)H(u)\rangle$ has poles at
$u\in\Z'_-=\{-\frac12,-\frac32,-\frac52,\dots\}$ and at
$v\in\Z'_+=\{\frac12,\frac32,\frac52,\dots\}$. This is readily
seen from the formula of Proposition 4.5.

3) Let us explain what we mean by application of the L'Hospital
rule. In the proof below we actually show that the diagonal
entries of the kernel are given by
$$
K(x,x)=\underset{u=x}\RES \langle G(u)\rangle, \qquad x\in\Z',
$$
where, by definition,
$$
G(u)=G(u,\la)=\left(\frac{\partial}{\partial
v}E(-v)H(u)\right)_{v=u}\,.
$$
On the other hand, the above expression for $K(x,y)$ makes sense
not only when $x,y$ are (distinct) points on the lattice $\Z'$
but also if $y\in\Z'_+$ and $x$ is a complex number with $\re
x>0$, or if $x\in\Z'_-$ and $y$ is a complex number with $\re
y<0$. (Indeed, this follows from the preceding comment.) Then we
can apply the L'Hospital rule to examine the limit values of
this extended kernel on the diagonal, and it is readily seen
that
$$
\langle G(y)\rangle=\lim_{x\to y} K(x,y), \quad y\in\Z'_-\,;
\qquad \langle G(x)\rangle=\lim_{y\to x} K(x,y), \quad
x\in\Z'_+\,.
$$

\begin{proof} Fix $m=1,2,\dots$ and assume that
$u_1,\dots,u_m,v_1,\dots,v_m$ are complex variables subject to
appropriate constraints on the argument and the modulus, as in
Corollary 4.4. This will ensure existence of the necessary
averages.

We start with the determinantal identity of Proposition 2.2,
which we rewrite as
\begin{multline*}
\left\langle\det\begin{pmatrix}
  \frac{E(-v_1)H(u_1)}{v_1-u_1} & \dots  &  \frac{E(-v_1)H(u_m)}{v_1-u_m}\\
  \hdotsfor{3}  \\
  \frac{E(-v_m)H(u_1)}{v_m-u_1} & \dots & \frac{E(-v_m)H(u_m)}{v_m-u_m} \\
\end{pmatrix}\right\rangle\\
=\det\begin{pmatrix}
  \frac{\langle E(-v_1)H(u_1)\rangle}{v_1-u_1} & \dots
  &  \frac{\langle E(-v_1)H(u_m)\rangle}{v_1-u_m}\\
  \hdotsfor{3}  \\
  \frac{\langle E(-v_m)H(u_1)\rangle}{v_m-u_1}
  & \dots & \frac{\langle E(-v_m)H(u_m)\rangle}{v_m-u_m}
\end{pmatrix}
\end{multline*}
To justify the passage from formal series to actual functions we
use the same trick as in the proof of Proposition 4.5.

We multiply both sides of this identity by the product
$\prod\limits_{i=1}^m(v_i-u_i)$ (in more detail, we multiply the
$i$th row of the matrix in the left--hand side or in the
right--hand side by $(v_i-u_i)$), then we differentiate with
respect to $v_1,\ldots, v_m$, and finally we specialize
$v_1=u_1,\ldots, v_m=u_m$.

The multiplication of the $i$th row by $(v_i-u_i)$ has the
following consequences: First, the same factor in the
denominator of the diagonal entry is cancelled. Second, when we
apply $\partial/\partial v_i$ to an off--diagonal entry, we only
have to differentiate this factor $(v_i-u_i)$, because $v_i-u_i$
vanishes after specialization $v_i=u_i$.

Using these observations and the notation $G(u)$ introduce above
we obtain the identity
\begin{multline*}
\left\langle \det\begin{pmatrix}
  G(u_1) & \frac{E(-u_1)H(u_2)}{u_1-u_2} & \dots & \frac{E(-u_1)H(u_m)}{u_1-u_m} \\
  \frac{E(-u_2)H(u_1)}{u_2-u_1} &  G(u_2) & \dots & \frac{E(-u_2)H(u_m)}{u_2-u_m} \\
  \hdotsfor{4}  \\
  \frac{E(-u_m)H(u_1)}{u_m-u_1} & \frac{E(-u_m)H(u_2)}{u_m-u_2}
  & \dots & G(u_m) \\
\end{pmatrix}\right\rangle\\
=\det\begin{pmatrix}
  \langle G(u_1)\rangle
  & \frac{\langle E(-u_1)H(u_2)\rangle}{u_1-u_2}
  & \dots
  & \frac{\langle E(-u_1)H(u_m)\rangle}{u_1-u_m} \\
  \frac{\langle E(-u_2)H(u_1)\rangle}{u_2-u_1}
  & \langle G(u_2)\rangle
  & \dots
  & \frac{\langle E(-u_2)H(u_m)\rangle}{u_2-u_m} \\
  \hdotsfor{4}  \\
  \frac{\langle E(-u_m)H(u_1)\rangle}{u_m-u_1}
  & \frac{\langle E(-u_m)H(u_2)\rangle}{u_m-u_2}
  & \dots
  &  \langle G(u_m)\rangle \\
\end{pmatrix}
\end{multline*}

Recall that $E(-u)H(u)\equiv1$. Using this fact we can simplify
the determinant in the left--hand side, which gives
\begin{multline}\label{45A}
\left\langle \det\begin{pmatrix}
  G(u_1) & \frac{1}{u_1-u_2} & \dots & \frac{1}{u_1-u_m} \\
  \frac{1}{u_2-u_1} &  G(u_2) & \dots & \frac{1}{u_2-u_m} \\
  \hdotsfor{4}  \\
  \frac{1}{u_m-u_1} & \frac{1}{u_m-u_2}
  & \dots & G(u_m) \\
\end{pmatrix}\right\rangle\\
=\det\begin{pmatrix}
  \langle G(u_1)\rangle
  & \frac{\langle E(-u_1)H(u_2)\rangle}{u_1-u_2}
  & \dots
  & \frac{\langle E(-u_1)H(u_m)\rangle}{u_1-u_m} \\
  \frac{\langle E(-u_2)H(u_1)\rangle}{u_2-u_1}
  & \langle G(u_2)\rangle
  & \dots
  & \frac{\langle E(-u_2)H(u_m)\rangle}{u_2-u_m} \\
  \hdotsfor{4}  \\
  \frac{\langle E(-u_m)H(u_1)\rangle}{u_m-u_1}
  & \frac{\langle E(-u_m)H(u_2)\rangle}{u_m-u_2}
  & \dots
  &  \langle G(u_m)\rangle \\
\end{pmatrix}
\end{multline}

Next, we remark that we may remove the constraints on variables
$u_i,v_i$ and regard the above formulas as an identity of
meromorphic functions. Indeed, the off--diagonal entries in
right--hand side are meromorphic functions by virtue of
Proposition 4.5, and the same property for the diagonal entries
$\langle G(u_i)\rangle$ is verified using the remark preceding
Corollary 4.8. As for the left--hand side, we expand the
determinant, apply averaging term--wise, and then use the same
remark to conclude that any quantity of the form $\langle
G(u_{i_1})\dots G(u_{i_k})\rangle$ with $i_1<\dots<i_k$ is
meromorphic.

Now  take the residues of  both sides of (\ref{45A}) at
$u_i=x_i$, where $i=1,\dots,m$ and the $x_i$'s are pairwise
distinct points of $\Z'$. In the left--hand side, only the
product of diagonal entries gives a nontrivial contribution.
Comparing with the formula of Proposition 4.8, we conclude that
the result in the left--hand side of (\ref{45A}) is equal to
\begin{equation}\label{45B}
\underset{u_1=x_1}{\RES}\dots \underset{u_m=x_m}{\RES}
F_m(u_1,\dots,u_m)=\sgn(x_1)\dots\sgn(x_m)\,\rho_m(x_1,\dots,x_m).
\end{equation}

To handle the right--hand side of (\ref{45A}) we may assume,
without loss of generality, that among the $x_i$'s, the first
$k$ numbers are positive while the last $l=m-k$ numbers are
negative. Then it is convenient to write the matrix in the
right--hand side as a $2\times2$ block matrix, according to
partition $m=k+l$. Taking into account the location of poles of
$\langle E(-u_i)H(u_j)\rangle$ (see comment 2 after the
statement of the theorem) we can take the residues inside the
matrix in an appropriate way. Namely, the matrix entries in
block $(1,1)$ are equipped with symbol
$\underset{u_j=x_j}{\RES}$; those in block $(1,2)$ are equipped
with symbols $\underset{u_i=x_i}{\RES}\,
\underset{u_j=x_j}{\RES}$; in block $(2,1)$ there are no
residues at all; and in block $(2,2)$ we use
$\underset{u_i=x_i}{\RES}$.

In our present notation, the sign in (\ref{45B}) is equal to
$(-1)^l$. Using this fact and comment 3 to the statement of the
theorem we finally obtained the desired determinantal expression
$$
\rho_m(x_1,\dots,x_m)=\det[K(x_i,x_j)],
$$
where the kernel is given by (\ref{45C}).
\end{proof}

\subsection{The discrete hypergeometric kernel} Let us introduce some notation.
Let $h(x)$ be the function on $\Z'=\Z'_+\sqcup\Z'_-$ given by
$$
h(x)=\begin{cases}\dfrac{(zz')^{1/4}\xi^{x/2}(1-\xi)^{\frac{z+z'}{2}}
\sqrt{(z+1)_{x-\tfrac12}(z'+1)_{x-\tfrac12}}}{\Gamma(x+\tfrac12)},
\qquad x\in\Z'_+,\\
\dfrac{(zz')^{1/4}\xi^{-x/2}(1-\xi)^{-\frac{z+z'}{2}}
\sqrt{(-z+1)_{-x-\tfrac12}(-z'+1)_{-x-\tfrac12}}}{\Gamma(-x+\tfrac12)},
\qquad x\in\Z'_-
\end{cases}
$$
and $m(u)$ be the $2\times 2$ matrix--valued function given by
\begin{multline}\label{46B}
m(u)=\begin{pmatrix} m_{11}(u) & m_{12}(u) \\ m_{21}(u) &
m_{22}(u)
\end{pmatrix}\\
=\begin{pmatrix}
  F(-z,-z',u+\tfrac12;\tfrac{\xi}{\xi-1})
  & \dfrac{\sqrt{zz'\xi}}{1-\xi}\;
  \dfrac{F(1+z,1+z',-u+\tfrac32;
  \tfrac{\xi}{\xi-1})}{-u+\tfrac12} \\
  \dfrac{-\sqrt{zz'\xi}}{1-\xi}\;
  \dfrac{F(1-z,1-z',u+\tfrac32;
  \tfrac{\xi}{\xi-1})}{u+\tfrac12}
  &  F(z,z',-u+\tfrac12;\tfrac{\xi}{\xi-1})
\end{pmatrix}
\end{multline}

We also write the kernel $K(x,y)$ in matrix form
$$
K(x,y)=\begin{pmatrix} K_{11}(x,y) & K_{12}(x,y) \\ K_{21}(x,y)
& K_{22}(x,y) \end{pmatrix}
$$
where $x>0, y>0$ in $K_{11}$; $x>0, y<0$ in $K_{12}$; $x<0, y>0$
in $K_{21}$; $x<0, y<0$ in $K_{22}$.

\begin{cor}
With the notation introduced above, the correlation kernel for
the point process on $\Z'$ corresponding to the measure
$M_{z,z',\xi}$ can be written in the form
\begin{multline}\label{46A}
\begin{pmatrix} K_{11}(x,y) & K_{12}(x,y) \\ K_{21}(x,y)
& K_{22}(x,y) \end{pmatrix} =h(x)h(y)\\
\times\begin{pmatrix}
\dfrac{-m_{11}(x)m_{21}(y)+m_{21}(x)m_{11}(y)}{x-y}
  &  \dfrac{m_{11}(x)m_{22}(y)-m_{21}(x)m_{12}(y)}{x-y} \\
   \dfrac{m_{22}(x)m_{11}(y)-m_{21}(y)m_{12}(x)}{x-y}
   &  \dfrac{-m_{22}(x)m_{12}(y)+m_{12}(x)m_{22}(y)}{x-y}
\end{pmatrix}
\end{multline}
where the indeterminacies of type $0/0$ on the diagonal are
removed by the L'Hospital rule.
\end{cor}

\begin{proof} First of all, it is worth noting that the kernel
written above differs from that of Theorem 4.9 by the
transformation  $K(x,y)\mapsto h(x)K(x,y)(h(y))^{-1}$, which
does not affect the correlation functions.

The claim of the corollary is obtained by direct computation of
the kernel of Theorem 4.9 using the explicit expression of
Proposition 4.5 and the knowledge of the residues of
$F(a,b;c;\zeta)$ (here $\zeta=\frac\xi{\xi-1}$) at points
$c=0,-1,\dots$:
$$
\underset{c=-n}{\RES} F(a,b;c;\zeta)=(-1)^n\zeta^{n+1}\,
\frac{(a)_{n+1}(b)_{n+1}}{n!(n+1)!}F(a+n+1,b+n+1;n+2;\zeta),
$$
see \cite{Er}, 2.8~(19).
\end{proof}

Note that the result of Corollary 4.10 agrees with the result
obtained in \cite{BO-hyper}. The kernel (\ref{46A}) is called
the {\em discrete hypergeometric kernel\/}.

\begin{rem} As was pointed out in Borodin \cite{B-IMRN}, Section 8,
the matrix $m$ appears in a discrete Riemann--Hilbert problem.
Namely, set
$$
w(x)=
\begin{pmatrix}
      0 & -h^2(x) \\
      0 & 0 \\
    \end{pmatrix}, \quad x\in\Z'_+; \qquad
w(x)= \begin{pmatrix}
     0 & 0 \\
     -h^2(x) & 0 \\
   \end{pmatrix}, \quad x\in\Z'_-\,.
$$
We are looking for a $2\times 2$ matrix--valued function
$m=m(u)$ with simple poles such that
\begin{enumerate}
    \item $m(u)
\;\mbox{is analytic}\;\mbox{in}\;\mathbb{C}\setminus
\mathbb{Z}'. $ \item $
\;\underset{u=x}{\RES}\,m(u)=\lim\limits_{u\rightarrow
x}\left(m(u)w(x)\right),\;\;x\in \mathbb{Z}'. $ \item
$\;m(u)\rightarrow 1\;\;\mbox{as}\;\; u\rightarrow \infty.$
\end{enumerate}
One can show that  this problem has a unique solution, which is
the matrix (\ref{46B}).

It is worth noting that the formula of Proposition 4.5 can also
be written in terms of $m(u)$:
\begin{equation}\label{46C}
\left\langle
E(-v)H(u)\right\rangle_{M_{z,z',\xi}}=m_{11}(v)m_{22}(u)-m_{21}(v)m_{12}(u).
\end{equation}
\end{rem}

The ``jump'' condition (2) allows one to quickly derive
(\ref{46A}) from (\ref{46C})

\begin{rem}  One should not think that the $z$-measures
and their degenerations exhaust all known examples of Giambelli
compatible measures on partitions. There exists a wider class of
(generally speaking, complex) Giambelli compatible measures,
which are constructed as follows.

Take any algebra homomorphism $\pi:\La\to \C$, denote $t:=-\pi(\bp_1)$ (where
$\bp_1$ is the first power sum), and for any $\xi\in\C$ with $|\xi|<1$ set
$$
M_{\pi,\xi}(\la)=(1-\xi)^t\cdot\frac{\pi(Fs_\la)\dim\lambda}{|\la|!}
\,(-\xi)^{|\la|}, \qquad \la\in\Y.
$$
Then $\sum_{\la\in\Y} M_{\pi,\xi}(\lambda)=1$, see
\cite{BO-mult}. In fact, the $z$-measures are special cases of
measures $M_{\pi,\xi}$, see \cite{BO-mult}, \S2. One can prove
that any $M_{\pi,\xi}$ (or the corresponding point process) is
Giambelli compatible. Some examples of positive measures
$M_{\pi,\xi}$, other than the $z$-measures, can be found in
\cite{BO-mult}, \S\S6.1--6.2.

\end{rem}

\section{ The  Whittaker kernel.}

In this section we discuss some Giambelli compatible point processes on a
continuous space, the punctured line $\R^*=\R\setminus\{0\}$. These processes
provide a solution to a problem of harmonic analysis on the infinite symmetric
group (see \cite{PartI}, \cite{BO-MRL}, \cite{BO-hyper}, \cite{Ol-surv}); they
are determined (in a certain precise sense) by the measures $M^{(n)}_{z,z'}$.
The correlation functions of these processes were first found in \cite{B-AA} by
rather heavy computations. Then a simpler derivation was obtained in
\cite{BO-hyper}; it relies on a scaling limit transition from lattice processes
corresponding to measures $M_{z,z',\xi}$, as $\xi$ approaches the critical
value $\xi=1$. Here we aim to demonstrate that using the Giambelli
compatibility property makes it possible to substantially simplify and clarify
the initial approach of \cite{B-AA}. Since two detailed proofs have already
been published, we only sketch the main steps of the argument (note that it is
quite similar to that of \S4). Some omitted technical details can be recovered
with the help of \cite{PartII}, \cite{B-AA}.

\subsection{The spaces $\Om$ and  $\wt\Om$}
Let $\R^\infty$ denote the direct product of countably many
copies of $\R$ equipped with the product topology. By
$\wt\Omega$ we denote the subspace of triples
$\om=(\al,\be,\de)\in\R^\infty\times\R^\infty\times\R$  such
that
$$
\alpha=(\alpha_1\geq\alpha_2\geq\ldots\geq 0),\quad
\beta=(\beta_1\geq\beta_2\geq\ldots\geq 0),\quad
\sum\limits_{i=1}^{\infty}\alpha_i+\sum\limits_{j=1}^{\infty}\beta_j\leq
\de.
$$
The space $\wt\Om$ is locally compact in the induced topology.
We will use it as a ``source'' space $S$. By definition, the
morphism $\phi$ of algebra $\La$ into the algebra of functions
on $\wt\Om$ is determined on the generators $\bp_k\in\La$ as
follows
\begin{equation}\label{51A}
\phi(\bp_1)(\om)\equiv\de, \qquad \phi(\bp_k)(\om)=
\sum\limits_{i=1}^\infty\al_i^k+(-1)^{k-1}\sum\limits_{j=1}^{\infty}\beta_j^k,
\quad k=2,3,\dots\,.
\end{equation}
The map $\phi$ is an embedding. To simplify the notation, given
$f\in\La$, we will abbreviate $f(\om)=\phi(f)(\om)$. One can
prove that the functions $f(\om)$ are continuous on $\wt\Om$.

Finally, we will also need the subspace $\Om:=\{\om\in\wt\Om\mid
\de=1\}$, which is called the {\em Thoma simplex\/}. Note that
$\Om$ is compact.

\subsection{The measures $P_{z,z'}$ and $\wt P_{z,z'}$} Fix
parameters $z,z'$ as in section 4. It is known that there exists
a unique probability measure $P_{z,z'}$ on the Thoma simplex
$\Om$, such that
\begin{equation}\label{52A}
\int_{\Om}s_\la(\om)P_{z,z'}(d\om)=\frac{M^{(|\la|)}_{z,z'}(\la)}{\dim\la}\,,
\qquad \forall\la\in\Y.
\end{equation}
Uniqueness of $P_{z,z'}$ follows from the fact that the image of $\La$ is dense
in the space of continuous functions on the compact space $\Om$. Existence is a
more deep claim; it follows from a general theory developed in \cite{KOO} (see
also \cite{Ol-surv}). Note that (\ref{52A}) can be viewed as an
infinite--dimensional moment problem: an unknown measure is characterized by
its ``moments'', which are indexed by $\la$'s. The measures $P_{z,z'}$ are
interesting because they govern the decomposition of certain natural
representations of the infinite symmetric group, see \cite{KOV},
\cite{Ol-surv}, and references therein.

For certain reasons explained in \cite{PartII}, \cite{B-AA}, \cite{PartIII} we
prefer to deal with a modification of $P_{z,z'}$. Consider the gamma
distribution on the positive half--line $\R_{>0}$ with parameter $zz'$:
$$
\gam_{zz'}(dr)=\frac1{\Ga(zz')}\,r^{zz'-1}\,e^{-r}dr, \qquad
r>0.
$$
The modified measure, denoted as $\wt P_{z,z'}$, lives on
$\wt\Om$ and is defined as the pushforward of
$P_{z,z'}\otimes\gam_{zz'}$ under the map
$$
\Om\times\R_{>0}\to\wt\Om, \qquad
((\al,\be),r)\mapsto(r\cdot\al,r\cdot\be,r).
$$
Clearly, $\wt P_{z,z'}$ is again a probability measure.

In a certain precise sense, $P_{z,z'}$ is the limit of measures
$M^{(n)}_{z,z'}$ as $n\to\infty$ while $\wt P_{z,z'}$ is the
limit of measures $M_{z,z',\xi}$ as $\xi\to1$.

\subsection{Giambelli compatibility}
It is readily verified that all functions $f(\om)$ on $\wt\Om$
coming from elements $f\in\La$ are integrable with respect to
$\wt P_{z,z'}$. Hence, the map $\phi$ as defined in \S5.1 sends
$\La$ to $\A(\wt\Om,\wt P_{z,z'})$.

\begin{prop}
The triple $(\wt\Om, \wt P_{z,z'}, \phi)$ is Giambelli
compatible.
\end{prop}

\begin{proof} Set $n=|\la|$ and observe that
$s_\la(r\cdot\om)=r^n\cdot s_\la(\om)$. It follows that
$$
\int_{\wt\Om}s_\la(\om)\wt P_{z,z'}(d\om)
=(zz')_n\,\int_{\Om}s_\la(\om)P_{z,z'}(d\om)
=(zz')_n\,\frac{M^{(n)}_{z,z'}(\la)}{\dim\la}\,.
$$
Denoting integration with respect to $\wt P_{z,z'}$ as
$\langle\,\cdot\,\rangle_{\wt P_{z,z'}}$ we thus get
$$
\langle s_\la\rangle_{\wt
P_{z,z'}}=(zz')_n\,\frac{M^{(n)}_{z,z'}(\la)}{\dim\la}\,.
$$
Then we use formula (\ref{41A}) and the expression of $\dim\la$
in terms of Frobenius coordinates, as in the proof of
Proposition 4.2.
\end{proof}

\subsection{Computation of $\left\langle E(v)H(u)\right\rangle_{P_{z,z'}}$
and  $\left\langle E(v)H(u)\right\rangle_{\wt P_{z,z'}} $} The
definition of $\phi$ (see (\ref{51A})) implies that
\begin{equation*}
H(u)(\om)=e^{\ga
u^{-1}}\,\prod\limits_{i=1}^{\infty}\frac{1+\beta_iu^{-1}}{1-\alpha_iu^{-1}},
\qquad E(v)(\om)=e^{\ga
v^{-1}}\,\prod\limits_{i=1}^{\infty}\frac{1+\alpha_iv^{-1}}{1-\beta_iv^{-1}},
\end{equation*}
where we are using the notation
$$
\ga=\de-\sum_i\al_i-\sum_j\be_j\,.
$$
Actually, $\ga=0$ almost surely (with respect to probability
measure $\wt P_{z,z'}$), see Theorem 6.1 in \cite{Ol-surv}.
Hence, the exponential prefactors could be omitted. However, a
priori we cannot use this fact because it appears as a
consequence of the computation of the correlation functions.

We will regard $u$ and $v$ as complex variables. Note that
infinite products are well defined provided that
$u,v\in\C\setminus\R$.

Lemma 4.3 implies that $\langle H(u)E(v)\rangle_{\wt P_{z,z'}}$
makes sense when $(u,v)$ ranges over a suitable domain in
$(\C\setminus\R)^2$. As in section 4, the possibility of
computing this quantity is based on the knowledge of $\langle
s_\la\rangle_{\wt P_{z,z'}}$. Our computation goes in two steps.
First we evaluate the average over $(\Om, P_{z,z'})$ and then we
pass to $\wt\Om$ using the ray integral transform with respect
to measure $\gam_{zz'}$. The reason is that on the Thoma simplex
$\Om$ we can use the formula
\begin{equation}\label{54A}
\langle H(u)E(v)\rangle=1+(u+v)\sum_{p,q\ge0}\frac{\langle
s_{(p\mid q)}\rangle}{u^{p+1}v^{q+1}}
\end{equation}
whereas on $\wt\Om$ such a series diverges.

Let $F_3(a, a', b, b';c; x,y)$ denote the hypergeometric
function in two variables $x$ and $y$, defined by the series
$$
F_3(a,a', b,b';c;
x,y)=\sum\limits_{m,n=0}^{\infty}\frac{(a)_m(a')_n(b)_m(b')_n}{(c)_{m+n}m!n!}
x^my^n.
$$
It possesses an Euler--type integral representation
\begin{multline*}
F_3(a,a',b,b';c;x,y)\\=
\frac{\Gamma(c)}{\Gamma(b)\Gamma(b')\Gamma(c-b-b')}
\iint\limits_{\substack{s\ge0,\, t\ge0
\\s+t\le1}}\frac{s^{b-1}t^{b'-1}(1-s-t)^{c-b-b'-1}ds
dt}{(1-sx)^{a}(1-ty)^{a'}}
\end{multline*}
(see Erdelyi \cite{Er}, \S5.7--5.8). The series converges in the
polydisc $|x|<1$, $|y|<1$ and can be analytically continued to a
larger domain using the integral representation. Note that using
the generalized function $s_+^a/\Ga(a)$ supported by the
half--line $s\ge0$ (see Gelfand and Shilov \cite{GS}), the
integral representation can be rewritten as
\begin{multline}\label{54B}
F_3(a,a',b,b';c;x,y)\\= \Gamma(c)\,
\iint\frac{s_+^{b-1}}{\Ga(b)} \frac{t_+^{b'-1}}{\Ga(b')}
\frac{(1-s-t)_+^{c-b-b'-1}}
{\Ga(c-b-b')}\frac{ds\,dt}{(1-sx)^{a}(1-ty)^{a'}}\,.
\end{multline}

\begin{lem}
For $u,v\in \C\setminus[0,1]$
\begin{multline*}
\left<E(v)H(u)\right>_{P_{z,z'}}=F_3(z,-z,z',-z';zz';u^{-1},v^{-1})\\
+\frac{1}{uv(zz'+1)}F_3(z+1,-z+1,z'+1,-z'+1;zz'+2;u^{-1},v^{-1}).
\end{multline*}
\end{lem}

\begin{proof} For $u,v\in \C\setminus[0,1]$,  $H(u)E(v)(\om)$ is
uniformly bounded on $\om\in\Om$, hence the quantity
$\left<E(v)H(u)\right>_{P_{z,z'}}$ is well defined and is a
holomorphic function in $u,v$. Assume that $|u^{-1}|<1$,
$|v^{-1}|<1$ first. Then we may apply formula (\ref{54A}), where
$\langle\,\cdot\,\rangle$ means
$\langle\,\cdot\,\rangle_{P_{z,z'}}$. Using the explicit
expression
$$
\langle s_{(p\mid
q)}\rangle_{P_{z,z'}}=\frac{M^{(p+q+1)}_{z,z'}((p\mid
q))}{\dim(p\mid q)} =
\frac{(z+1)_p(z'+1)_p(-z+1)_q(-z'+1)_q}{(zz'+1)_{p+q}p!q!(p+q+1)}
$$
one can verify the desired formula directly. Then we use
analytic continuation.
\end{proof}

This completes the first step. The second step, the passage to
average over $\wt\Om$, is based on the relation

\begin{equation*}
\left\langle H(u)E(v)\right\rangle_{\wt
P_{z,z'}}=\frac{1}{\Gamma(zz')}\int\limits_0^{\infty} \langle
H(ur^{-1})E(vr^{-1})\rangle_{P_{z,z'}}\,r^{zz'-1}e^{-r}dr.
\end{equation*}

It turns out that the result is expressed through the classical
Whittaker function $W_{\ka,\mu}(x)$ (see \cite{Er}, \S6, for the
definition). This function possesses the integral representation
(see \cite{Er}, 6.11 (18))
\begin{equation}\label{54C}
W_{\ka,\mu}(x)=e^{-x/2}x^{\mu+1/2}\int_0^\infty
\frac{t_+^{\ka+\mu-1/2}}{\Ga(-\ka+\mu+1/2)}(1+t)^{\ka+\mu+1/2}e^{-xt}dt.
\end{equation}
The integral converges for $\re x>0$ and admits an analytic
continuation to the larger domain $\C\setminus(-\infty,0]$.

In the next proposition we assume that $u,v\in\C\setminus\R$ are
such that $(H(u)E(v))(\om)$ is integrable with respect to
measure $\wt P_{z,z'}$ on $\wt\Om$. By Lemma 4.3, this holds at
least for for large $|u|$ and $|v|$.

\begin{prop}
Under these assumptions we have
\begin{multline*}
\langle H(u) E(v)\rangle_{\wt P_{z,z'}}\\=e^{-\frac{v+u}{2}}
 \left((-v)^{-\frac{z+z'+1}{2}}
W_{\frac{z+z'+1}{2},\frac{z-z'}{2}}(-v)\cdot(-u)^{\frac{z+z'-1}{2}}
W_{\frac{-z-z'+1}{2},\frac{z-z'}{2}}(-u)\right.
\\
+\left.zz'(-v)^{-\frac{z+z'+1}{2}}
W_{\frac{z+z'-1}{2},\frac{z-z'}{2}}(-v)\cdot
(-u)^{\frac{z+z'-1}{2}}W_{-\frac{z+z'+1}{2},\frac{z-z'}{2}}(-u)\right).
\end{multline*}
\end{prop}

\begin{proof} Direct computation using Lemma 5.2 and integral
representations  (\ref{54B}) and (\ref{54C}).
\end{proof}

\begin{rem} On a heuristical level, this result can be obtained
directly from (\ref{54A}) with $\langle\,\cdot\,\rangle$
understood as $\langle\,\cdot\,\rangle_{\wt P_{z,z'}}$. The
formal summation leads to
\begin{multline*}
_{2}F_{0}(z,z';u^{-1})\,\,
_{2}F_{0}(-z,-z';v^{-1})\\
+\frac{zz'}{uv}\,\, _{2}F_{0}(z+1,z'+1;u^{-1})\,\,
_{2}F_{0}(-z+1,-z'+1;v^{-1}).
\end{multline*}
Here $_2F_0(a,b;x)$ is a divergent hypergeometric series, which,
however, can be interpreted as an asymptotic series for the
Whittaker function (see \cite{Er}, 6.9 (5)).
\end{rem}

\subsection{Correlation measures and controlling measures}
The contents of the present subsection is similar to that of
\S4.4.

Instead of the lattice $\Z'$ we are dealing with punctured line
$\R^*=\R\setminus\{0\}$. To an arbitrary point
$\om=(\al,\be,\de)\in\wt\Om$ we assign a point configuration
$X(\om)\subset\R^*$ as follows: we remove the possible 0's from
the sequences $\al$ and $\be$ and then set
$$
X(\om)=\{-\be_1,-\be_2,\dots,\al_2,\al_1\}.
$$
For instance, in the special case when both $\al$ and $\be$ are
zero sequences, the configuration $X(\om)$ is empty. Note that
the correspondence $\om\mapsto X(\om)$ is not injective, because
we cannot restore $\de$ from $X(\la)$. However, the restriction
to the subset of $\om$'s with $\ga=0$ is injective.

Assume we are given a probability measure $P$ on $\wt\Om$. Then
we obtain a point process on $\R^*$ with ``source space''
$(\wt\Om, P)$. Let $\rho_m$ stand for the $m$th correlation
measure of this process. As a reference measure on $\R^*$ we
take Lebesgue measure. If $\rho_m$ is absolutely continuous with
respect to Lebesgue measure, we can pass to the correlation
function, which we will denote as $\rho_m(x_1,\dots,x_m)$. (Even
if $\rho_m$ is not absolutely continuous,
$\rho_m(x_1,\dots,x_m)$ makes sense as a generalized function.)
Informally, $\rho_m(x_1,\dots,x_m)$ is the density of the
probability that the random configuration intersects each of the
infinitesimal intervals $[x_i,x_i+dx_i]$, $i=1,\dots,m$.

Next, we assign to any $\om\in\wt\Om$ a measure on $\R$,
$$
\si_\om=\sum_{i=1}^\infty(\al_i\de_{\al_i}+\be_i\de_{-\be_i})
+\ga\de_0,
$$
of total mass $\de$, and then we define the $m$th controlling
measure $\si_m$ on $\R^m$ ($m=1,2,\dots$) as follows:
$$
\si_m=\int_{\wt\Om}\si_\om^{\otimes m} P(d\om).
$$
The controlling measures contain all the information about the
correlation measures, see \cite{PartI}. In particular, there is
a simple correspondence between the restrictions of $\si_m$ and
$\rho_m$ to the subset $(\R^*)^m_0\subset(\R^*)^m$ of vectors
$(x_1,\dots,x_m)$  with distinct coordinates:
$$
\si_m=|x_1\dots x_m|\,\rho_m \qquad \text{on $(\R^*)^m_0$}.
$$

Assuming that $\si_m$ satisfies the growth condition (\ref{44A})
we can introduce its Cauchy transform $\widehat\si_m$. It is
well known (and readily verified) that $\si_m$ can be restored
from $\widehat\si_m$ as follows
$$
\si_m(x)=\underset{u=x}\JUMP\, \widehat\si_m(u):= \frac1{2\pi
i}\,\lim_{\eps\downarrow0}(\widehat\si_m(x-i\eps)-\widehat\si_m(x+i\eps)),
\qquad x\in\R^m,
$$
where the limit means weak limit of generalized functions.

Arguing as in Lemma 4.7 we have
\begin{multline*}
\widehat\si_m(u_1,\dots,u_m)\\=u_1\dots
u_m\left\{\frac{\partial^m}{\partial v_1\dots\partial
v_m}\langle H(u_1)E(-v_1)\dots
H(u_m)E(-v_m)\rangle_P\right\}_{\substack{v_1=u_1\\
\dots\\ v_m=u_m}}
\end{multline*}
As in \S4, set
$$
F_m(u_1,\dots,u_m)=\left\{\frac{\partial^m}{\partial
v_1\dots\partial v_m}\langle H(u_1)E(-v_1)\dots
H(u_m)E(-v_m)\rangle_P\right\}_{\substack{v_1=u_1\\
\dots\\ v_m=u_m}}
$$
Then we have on $(\R^*)^m_0$
$$
\rho_m(x_1,\dots,x_m)=\sgn(x_1)\dots\sgn(x_m)\,\underset{u_1=x_1}{\JUMP}\dots
\underset{u_m=x_m}{\JUMP}\,\bigl[ F_m(u_1,\dots,u_m)\bigr].
$$

\subsection{Computation of the correlation functions} We set $P=\wt P_{z,z}$.
One can verify that the corresponding controlling measures
$\si_m$ are finite measures, so that their Cauchy transforms
$\widehat\si_m$ are well defined.

\begin{thm} The point process on $\R^*$ corresponding to the measure $P_{z,z'}$ is
determinantal and its correlation kernel can be  written as
\begin{equation*}
K(x,y)=\left\{
\begin{array}{ll}
   \underset{u=y}{\JUMP}\;\dfrac{\langle E(-x)H(u)\rangle}{x-y}, & x>0, y>0,\; x\neq y, \\
   & \\
    -\underset{u'=x}{\JUMP}\;\underset{u=y}{\JUMP}\;
    \dfrac{\langle E(-u')H(u)\rangle}{x-y}, &  x<0, y>0,\\
    & \\
    \dfrac{\langle E(-x)H(y)\rangle}{x-y}, & x>0, y<0, \\
    & \\
   -\underset{u=x}{\JUMP}\;\dfrac{\langle E(-u)H(y)\rangle}{x-y}, & x<0, y<0,\; x\neq y,  \\
\end{array}
\right.
\end{equation*}
where $\langle\,\cdot\,\rangle$ means
$\langle\,\cdot\,\rangle_{\wt P_{z,z'}}$, and the indeterminacy
arising for $x=y$ is resolved via the L'Hospital rule.
\end{thm}

\begin{proof}[Idea of proof]
We compute $\rho_m$ on the subset $(\R^*)^m_0$ of $(\R^*)^m$
(one can check that this subset has full measure with respect to
$\rho_m$, see Theorem 2.5.1 in \cite{PartII}). The scheme of the
argument is similar to that of the proof of Theorem 4.9: we use
the formula of Proposition 5.3 and and the determinantal
identity (\ref{2B}). Let us briefly describe how to justify this
identity. Here we cannot apply the trick of Proposition 4.5;
instead of this we rearrange the proof of Proposition 2.2 using
the two--step procedure of \S5.4. Namely, we start with
integration over the Thoma simplex:
$$
\left\langle\det\left(\frac{H(u_i)E(v_j)-
1}{u_i+v_j}\right)_{i,j=1}^d\right\rangle_{P_{z,z'}}
={\sum\limits_{\substack {p_1,\ldots ,\; p_d=0\\
{q_1,\ldots
,\;q_d=0}}}^{\infty}}\frac{\langle\det\left(s_{(p_i|q_j)}\right)_{i,j=
1}^d\rangle_{P_{z,z'}}} {u_1^{p_1+1}\ldots
u_d^{p_d+1}v_1^{q_1+1}\ldots v_d^{q_d+1}}
$$

Then, using the explicit expression of Proposition 5.3 we write
the sum in terms of an integral:
\begin{multline*}
\frac{\Gamma(zz') (zz')^d}{u_1\dots u_d\,v_1\dots v_d} \int\limits_{\substack{s_i,t_i>0\\
i=1,\dots,d}} \,\int\limits_{\substack{w_i\in(0,1)\\
i=1,\dots,d}} \sum_{\tau\in S_d}\sgn(\tau)\\
\times\prod_{i=1}^d
\frac{s_i^{z'}}{\Gamma(z'+1)}\frac{t_i^{-z'}}{\Gamma(-z'+1)}\left(1-\frac{w_i
s_i}{u_i}\right)^{-z-1}\left(1-\frac{w_{\tau(i)}
t_i}{v_i}\right)^{z-1}\\
\times
\frac{(1-\sum_{i=1}^d(s_i+t_i))^{zz'-d-1}_+}{\Gamma(zz'-d)}\,\,dw_i\,ds_idt_i.
\end{multline*}
where $S_d$ is the symmetric group of degree $d$ and
$\sgn(\tau)$ stands for the signature of a permutation $\tau\in
S_d$. To see the equivalence it suffices to expand the factors
$(1-...)^{\pm z-1}$ and use the Dirichlet integral. The
integration over $w_i$'s can be explicitly performed, see proof
of Lemma 2.2.4 in \cite{PartII}, which simplifies the formula.
In particular, the sum over $\tau$ can be turned into a
determinant inside the integral.

This formula splits into a $d\times d$ determinant of
``2-point'' averages $\langle\,\cdot\,\rangle_{\wt P_{z,z'}}$
under the ray transform, which follows from the one-dimensional
integration formula
$$
\int_0^{\infty}
\frac{(1-r^{-1}\sum_{i=1}^d(s_i+t_i))^{zz'-d-1}_+}
{\Gamma(zz'-d)}\,\,r^{zz'-d-1}e^{-r}dr=e^{-\sum_i(s_i+t_i)}.
$$
\end{proof}

\subsection{The Whittaker kernel}
Write $\R^*$ as $\R_+\sqcup\R_-$ (strictly positive and strictly
negative reals) and define a function $h$ on $\R^*$ by
$$
h(x)=\begin{cases}
\dfrac{(zz')^{1/4}}{\sqrt{\Gamma(z+1)\Gamma(z'+1)}}x^{(z+z')/2}e^{-x/2},
\quad x>0\\
\dfrac{(zz')^{1/4}}{\sqrt{\Gamma(-z+1)\Gamma(-z'+1)}}(-x)^{-(z+z')/2}e^{x/2},
\quad x<0.
\end{cases}
$$
Let $m(u)$ be the following $2\times2$ matrix--valued function
on $\C\setminus\R$:
$$
\begin{pmatrix} u^{-\frac{z+z'+1}{2}}e^{\frac{u}{2}}
W_{\frac{z+z'+1}{2},\frac{z-z'}{2}}(u) &
\sqrt{zz'}\,(-u)^{\frac{z+z'-1}{2}}
e^{-\frac{u}{2}}W_{\frac{-z-z'-1}{2},\frac{z-z'}{2}}(-u) \\
 -\sqrt{zz'}\, u^{-\frac{z+z'+1}{2}}e^{\frac{u}{2}}
 W_{\frac{z+z'-1}{2},\frac{z-z'}{2}}(u)
 & (-u)^{\frac{z+z'-1}{2}}e^{-\frac{u}{2}}
 W_{\frac{-z-z'+1}{2},\frac{z-z'}{2}}(-u)
\end{pmatrix}
$$

\begin{cor}
After the transformation $K(x,y)\to h(x)K(x,y)(h(y))^{-1}$ the
kernel of Theorem 5.5 can be written in the same form as in
Corollary 4.10, with $h$ and $m$ as defined above.
\end{cor}

This is exactly the {\em Whittaker kernel\/} of \cite{B-AA}, \cite{BO-hyper}.
About this kernel, see also \cite{BOk}, \cite{PartV}.

As in \S4, $m(u)$ turns out to be a solution to a
Riemann--Hilbert problem, see \cite{B-IMRN}.

Finally, note that the expression of Proposition 5.3 can be
written in terms of $m(u)$, just as in (\ref{46C}).


\begin{thebibliography}{99}

\bibitem{AP} G.~Akemann, A.~Pottier, Ratios of characteristic polynomials in complex
matrix models, J.Phys. A37 (2004) L453-L460; math-ph/0404068

\bibitem{AS}
A.~V.~Andreev, B.~D.~Simons,. Correlators of spectral
determinants in Quantum Chaos. {\em Phys. Rev. Lett.}
\textbf{75} (12), (1995) 2304-2307.

\bibitem{BDS}
J.~Baik, P~Deift, and E.~Strahov, Products and ratios of
characteristic polynomials of random hermitian matrices. {\em J.
Math. Phys}. \textbf{44}, (2003) 3657-3670.

\bibitem{BG}
D.~Bump and A.~Gamburd, On the averages of characteristic
polynomials from classical groups, math-ph/0502043

\bibitem{B-IMRN}
A.~Borodin, Riemann-Hilbert Problem and the Discrete Bessel
Kernel,  Intern. Math. Research Notices, 2000, no.~9, 467--494.

\bibitem{B-AA}
A.~M.~Borodin, Harmonic analysis on the infinite symmetric
group, and the Whittaker kernel,  Algebra i Analiz 12 (2000),
no. 5, 28-63 (Russian); English translation in St.~Petersburg
Math. J. 12 (2001), no.~5, 733--759.

\bibitem{PartII}
A.~Borodin,  Point processes and the infinite symmetric group.
Part II: Higher correlation functions, math/9804087.

\bibitem{BOk}
A.~Borodin and A.~Okounkov, A Fredholm determinant formula for
Toeplitz determinants, Integral Equations and Operator Theory 37
(2000) 386--396; math/9907165.

\bibitem{BOO}
A.~Borodin, A.~Okounkov, and G.~Olshanski, Asymptotics of
Plansherel measures for symmetric groups. J. Amer. Math. Soc.
13 (2000), 491--515.

\bibitem{BO-MRL}
A.~Borodin and G.~Olshanski, Point processes and the infinite
symmetric group. Math. Res. Lett. 5 (1998), 799--816.

\bibitem{PartIII}
A.~Borodin and G.~Olshanski, Point processes and the infinite
symmetric group. Part III: Fermion point processes,
math/9804088.

\bibitem{BO-hyper}
A.~Borodin and G.~Olshanski, Distributions on partitions, point
processes and the hypergeometric kernel.  Commun. Math. Phys.
211 (2000), 335--358.

\bibitem{BO-mult}
A.~Borodin and G.~Olshanski, Harmonic functions on
multiplicative graphs and interpolation polynomials, Electron.
J. Combin. 7 (2000) paper \#R28; math/9912124

\bibitem{BO-RSK}
A.~Borodin and G.~Olshanski, Z-Measures on partitions,
Robinson--Schensted--Knuth correspondence, and $\beta=2$
ensembles. In:  Random matrix models and their applications
(P.~M.~Bleher and A.~R.~Its, eds.). MSRI Publications, vol. 40,
Cambridge Univ. Press, 2001, pp. 71--94.

\bibitem{BO-Uinfty}
A.~Borodin and G.~Olshanski, Harmonic analysis on the infinite--dimensional
unitary group and determinantal point processes. Annals of Mathematics, 161
(2005), 1319--1422.

\bibitem{BO-Gamma}
A.~Borodin and G.~Olshanski, Random Partitions and the gamma kernel,  Advances
in Math. 194 (2005), 141--202.

\bibitem{BO-dynam}
A.~Borodin and G.~Olshanski, Markov processes on partitions,
math-ph/0409075.

\bibitem{BS}
A.~Borodin and E.~Strahov, Averages of characteristic polynomials in Random
Matrix Theory.  Commun. Pure and Applied Math. 59 (2006), no. 2, 161--253.

\bibitem{BH1}
E.~Brezin, S.~Hikami, Characteristic polynomials of random
matrices. Commun. Math. Phys. 214 (2000), 111--135.

\bibitem{BH2}
E.~Brezin, S.~Hikami, Characteristic polynomials of random
matrices at edge singularities. Phys. Rev. E 62 (2000)
3558--3567.

\bibitem{BH3}
E.~Brezin, S.~Hikami, Characteristic polynomials of real
symmetric random matrices. Commun. Math. Phys. 223 (2001)
363--382.

\bibitem{BH4}
E.~Brezin, S.~Hikami, New correlation functions for random
matrices and integrals over supergroups. J. Phys. A. 36 (2003)
711--751.

\bibitem{CFKRS}
J.~B.~Conrey, D.~W.~Farmer, J.~P.~Keating, M.~O.~Rubinshtein,
N.~C.~Snaith,  Autocorrelation of random matrix polynomials.
Commun. Math. Phys. 237 (2003), 365--395.

\bibitem{CFS}
J.~B.~Conrey, P.~J.~Forrester, and N.~C.~Snaith, Averages of ratios of
characteristic polynomials for the compact classical groups, Intern. Math. Res.
Notices (2005), no. 7, 397--431.

\bibitem{DVJ}
D.~J.~Daley, D.~Vere-Jones, An introduction to the theory of
point processes. Springer series in statistics, Springer, 1988.

\bibitem{De}
P.~Deift,  Orthogonal Polynomials and Random Matrices: A
Riemann-Hilbert Approach. Courant lecture notes, 3.  New York:
Courant Institute of Mathematical sciences, New York University
2000.

\bibitem{Er}
A.~Erdelyi (ed.), Higher transcendental functions. Vol. 1,
McGraw-Hill, (1953)


\bibitem{Fyo}
Y.~V.~Fyodorov,  Complexity of Random Energy Landscapes, Glass Transition and
Absolute Value of Spectral Determinant of Random Matrices. Phys. Rev. Lett. 92
(2004), paper \#240601.

\bibitem{FS}
Y.~V.~Fyodorov and E.~Strahov, An exact formula for general
spectral correlation function of random Hermitian matrices. J.
Phys. A 36 (2003) 3203--3213.

\bibitem{GS}
I.~M.~Gelfand and G.~E.~Shilov, Generalized functions:
properties and operations. Academic Press, New York, 1964.

\bibitem{GGK}
J.~Gronqvist, T.~Guhr, H.~Kohler,  The k-point random matrix
kernels obtained from one-point supermatrix models.  J. Phys. A.
37 (2004) 2331--2344.

\bibitem{GTW}
J.~Gravner, C.~A.~Tracy, and H.~Widom, Limit Theorems for Height
Fluctuations in a Class of Discrete Space and Time Growth
Models, J. of Statistical Physics 102 (2001), 1085--1132.

\bibitem{HKO}
C.~P.~Hughes, J.~P.~Keating, and N.~O'Connell, On the
characteristic polynomial of a random unitary matrix.  Commun.
Math. Phys. 220 (2001) 429--451.

\bibitem{Jo-shape}
K.~Johansson, Shape fluctuations and random matrices, Commun.
Math. Phys.  209(2000), 437--476; math/9903134.

\bibitem{Jo-planch}
K.~Johansson, Discrete orthogonal polynomial ensembles and the
Plancherel measure,  Ann. Math.  153 (2001), 259--296;
math/9906120.

\bibitem{Jo-nonint}
K.~Johansson, Non--intersecting paths, random tilings and random
matrices. Probab. Theory and Related Fields 123 (2002),
225--280.


\bibitem{Jo-ECM}
K.~Johansson, Random growth and random matrices, In: 3rd
European congress of mathematics (Barcelona, Spain, July 10-14,
2000). Volume I. Progr. Math. 201.  Birkh\"auser, Basel, 2001,
pp. 445-456.

\bibitem{KS1}
J.~P.~Keating, N.~C.~Snaith, Random Matrix Theory and
zeta(1/2+it). Commun. Math. Phys. 214 (2000), 57--89.

\bibitem{KS2}
J.~P.~Keating, N.~C.~Snaith, Random matrices and L-functions. J.
Phys. A. 36 (2003), no. 12, 2859--2881.

\bibitem{KS3} J.~P.~Keating, N.~C.~Snaith,
Random matrix theory and L-functions at s=1/2, Commun. Math.
Phys. 214 (2000), 91--110.

\bibitem{Ke}
S.~V.~Kerov, Anisotropic Young diagrams and Jack symmetric
functions, Funktsional. Anal. Prilozhen. 34 (2000), no.~1,
51--64 (Russian); English translation: Funct. Anal. Appl., 34
(2000),  41--51; math/9712267.

\bibitem{KOO}
S.~Kerov, A.~Okounkov, and G.~Olshanski, The boundary of the
Young graph with Jack edge multiplicities. Intern. Math.
Research Notices, 1998, no. 4, 173--199.

\bibitem{KO}
S.~Kerov and G.~Olshanski, Polynomial functions on the set of
Young diagrams, Comptes Rendus Acad. Sci.Paris S\'er. I, 319
(1994), 121--126.

\bibitem{KOV}
Kerov, S., Olshanski, G., and Vershik, A.: Harmonic analysis on
the infinite symmetric group. {\em Invent. Math.} \textbf{158}
(2004), no. 3, 551-642.

\bibitem{KOR}
W.~K\"onig,  N.~O'Connell, and S.~Roch,  Non--colliding random
walks, tandem queues, and discrete orthogonal polynomial
ensembles, Electronic J. Prob. Vol. 7 (2002) Paper No. 1, pages
1--24.

\bibitem{Le}
A.~Lenard, Correlation functions and the uniqueness of the state
in classical statistical mechanics, Commun. Math. Phys, 30
(1973), 35--44.

\bibitem{Ma}
 I., G. Macdonald, Symmetric functions and Hall polynomials, 2nd
edition, Oxford University Press, 1995.

\bibitem{Me}
M.~L.~Mehta, Random Matrices, 2nd ed., Academic Press, San
Diego, 1991.

\bibitem{O'C}
N.~O'Connell, Conditioned random walks and the RSK
correspondence, J. Phys. A: Math. Gen. 36 (2003) 3049-3066.

\bibitem{Ok-SL2}
A.~Okounkov, $SL(2)$ and z--measures.  In: Random matrix models
and their applications (P.~M.~Bleher and A.~R.~Its, eds). MSRI
Publications, vol. 40. Cambridge Univ. Press, 2001, pp. 71--94;
math/0002135.

\bibitem{Ok-infw}
A.~Okounkov, Infinite wedge and measures on partitions, Selecta
Math. (New Series), 7 (2001), 1--25; math/9907127.

\bibitem{PartV}
G.~Olshanski, Point processes and the infinite symmetric group.
Part V: Analysis of the matrix Whittaker kernel, math/98140014.

\bibitem{PartI}
G.~Olshanski, Point processes related to the infinite symmetric
group. In: The orbit method in geometry and physics: in honor of
A.~A.~Kirillov (Ch.~Duval, L.~Guieu, and V.~Ovsienko, eds.),
Progress in Mathematics 213, Birkh\"auser, 2003, pp. 349--393.

\bibitem{Ol-surv}
G.~Olshanski, An introduction to harmonic analysis on the
infinite symmetric group. In: Asymptotic combinatorics with
applications to mathematical physics (St. Petersburg, 2001),
Lecture Notes in Math., 1815, Springer, Berlin, 2003, 127--160.

\bibitem{ORV}
G.~Olshanski, A.~Regev, A.~Vershik, Frobenius-Schur functions.
In: Studies in Memory of Issai Schur (A.~Joseph, A.~Melnikov,
and R.~Rentschler, eds.), Progress in Mathematics 210,
Birkh\"auser, 2003, pp. 251--300.; math.CO/0110077.

\bibitem{Ra}
E.~M.~Rains, Correlation functions for symmetrized increasing
subsequences, math/0006097.

\bibitem{SF}
E.~Strahov and Y.~V.~Fyodorov, Universal Results for
Correlations of Characteristic Polynomials: Riemann-Hilbert
Approach. Commun. Math. Phys. 241 (2003) 343-382.

\bibitem{TW}
C.~A.~Tracy and H.~Widom, Correlation Functions, Cluster
Functions and Spacing Distributions for Random Matrices. J.
Stat. Phys. 92 (1998), 809--835.

\bibitem{VK}
A.~M.~Vershik and S.~V.~Kerov, Asymptotic theory of characters
of the symmetric group, Funktsion. Anal. Prilozhen. 15 (1981),
no.~4, 15--27 (Russian); English translation: Funct. Anal.
Appl., 15 (1981), 246--255.

\end{thebibliography}
\end{document}